\documentclass[aps,prb,twocolumn,superscriptaddress,longbibliography,footinbib]{revtex4-1}  %
\usepackage{graphicx} 
\usepackage{float} 
\usepackage{dcolumn} 
\usepackage{bm,color}
\usepackage{amsmath,amssymb,dsfont,amstext,amsfonts}
\usepackage{extarrows}
\usepackage{xcolor}
\usepackage{wasysym}
\usepackage{mathtools}
\usepackage{bbold}
\usepackage[export]{adjustbox}
\usepackage{mathdots}
\usepackage{siunitx}
\usepackage{comment}
\usepackage[colorlinks=true,linkcolor=blue,urlcolor=blue,citecolor=blue]{hyperref}

\usepackage[normalem]{ulem} 
\usepackage{color}

\begin{document}
\title{Correlation-driven phonon renormalisation and the equation of state of  $\gamma$-cerium}

\author{Yao Wei}
\email{yao.wei@kcl.ac.uk}
\affiliation{Theory and Simulation of Condensed Matter
(TSCM), King's College London, Strand, London WC2R 2LS, United Kingdom}
\affiliation{Institute of Physics (FZU), Czech Academy of Sciences, Na Slovance 2, 182 00 Praha, Czech Republic}
\affiliation{Institute of Informatics, Slovak Academy of Sciences, 845 07 Bratislava, Slovakia}
\author{Siyu Chen}
\affiliation{Cavendish Laboratory, University of Cambridge, Cambridge CB3 0HE,  United Kingdom}
\affiliation{Department of Materials Science and Metallurgy, University of Cambridge, Cambridge CB3 0FS,  United Kingdom}
\author{Evgeny Plekhanov}
\affiliation{Theory and Simulation of Condensed Matter
(TSCM), King's College London, Strand, London WC2R 2LS, United Kingdom}
\author{Ivan Štich}
\affiliation{Institute of Informatics, Slovak Academy of Sciences, 845 07 Bratislava, Slovakia}
\author{Cedric Weber}
\email{cedric.weber.phd@gmail.com}
\affiliation{Quantum Brilliance Pty, The Australian National University, Canberra, Australian Capital Territory 2600, Australia}
\author{Jan M. Tomczak}
\email{jan.tomczak@kcl.ac.uk}
\affiliation{Theory and Simulation of Condensed Matter
(TSCM), King's College London, Strand, London WC2R 2LS, United Kingdom}
\affiliation{Institute of Solid State Physics, TU Wien, Vienna, Austria}

\begin{abstract}
We investigate the thermodynamic properties of elemental cerium by assessing the crucial role of phonon free energy within the framework of dynamical mean-field theory (DMFT). While conventional density functional theory (DFT) often fails to capture the intricate energy landscape of $f$-electron materials, our approach integrates many-body electronic correlations with lattice dynamics to achieve a more rigorous description of the equation of state. We calculate the total energy as a function of the lattice constant at both the DFT and DFT+DMFT levels, subsequently incorporating the vibrational free energy derived from the phonon density of states. Our findings reveal that electronic renormalisation of the force constants significantly alters the phonon spectra, particularly in the strongly correlated $\gamma$-phase. 
By applying these phonon corrections to the energy profiles, we observe a substantial refinement in the predicted equilibrium volumes.
Using principal-component-based machine learning, we interpolate phonon dispersions continuously from a finite set of first-principles calculations and compare them to experiment, finding significantly closer agreement compared to conventional DFT and DFT+U calculations that neglect dynamical many-body correlations. 
This study underlines the necessity of accounting for both electronic and vibrational entropy when evaluating the phase stability and structural transitions of lanthanide systems under varying pressures and temperatures. 
\end{abstract}
\maketitle

\section{Introduction}

Elemental cerium exhibits a remarkable isostructural $\gamma$→$\alpha$ phase transition with a volume collapse of approximately 15\%
\cite{koskenmaki1978cerium,lawson1949concerning}.
Spectroscopic measurements reveal a pronounced change in the $4f$ electronic character across
the transition, from incoherent, fluctuating local moments in the $\gamma$ phase, to a regime with coherent quasiparticles
in the $\alpha$ phase\cite{wieliczka1982photoemission,laubschat1990surface}.
Early theoretical proposals, therefore, centred on the nature of the $4f$ electrons, proposing a
Mott-like localisation transition driven by on-site Coulomb repulsion
\cite{johansson1974alpha}, or the Kondo volume collapse scenario, in which the quenching of local moments by conduction electrons generates the volume change
\cite{allen1982kondo,lavagna1982volume}.
A natural theoretical framework to understand the $\gamma$→$\alpha$ transition is
dynamical mean-field theory (DMFT) \cite{georges1996dynamical,kotliar2004strongly}.
In combination with density-functional theory, DFT+DMFT has demonstrated the volume
collapse and the associated spectroscopic and magnetic properties in cerium from first principles
\cite{held2001cerium,mcmahan2003thermodynamic,haule2005alpha,amadon2006alpha,chakrabarti2014alpha}.

Progress in the DFT+DMFT methodology to include correlated-electron force calculations
\cite{leonov2014first,haule2016forces,plekhanov2018many},
has further enabled increasingly quantitative thermodynamic and structural descriptions of the
$\gamma$→$\alpha$ transition.
Recent studies have successively incorporated additional contributions to the free energy in cerium,
moving from the electronic ground-state energy to the {\it electronic} entropy\cite{haule2016forces}, confirming previous suggestions\cite{amadon2006alpha} that the $\gamma$-phase is stabilized by entropy.

A key remaining question is the quantitative role of {\it vibrational} free energy in determining
The equilibrium lattice constants of both phases.
Experimental evidence on this point is not yet settled: inelastic neutron scattering on cerium alloys find small differences in the phonon density of states across the transition
\cite{manley2002vibrational}, while high-pressure diffraction studies indicate a vibrational
entropy change of order $0.75\,k_\mathrm{B}$/atom, roughly half the total entropy
change \cite{jeong2004role}.
Inelastic X-ray scattering measurements of the complete phonon dispersion of elemental cerium across the transition find an intermediate value of
$\sim 0.33\,k_\mathrm{B}$/atom
\cite{krisch2011phonons}, indicating a non-negligible but sub-dominant lattice
contribution.
Irrespective of its role in the transition itself, vibrational entropy is well known to
expand equilibrium volumes through the Gr\"{u}neisen mechanism, so that a fully
predictive calculation of the lattice constants of both $\alpha$- and $\gamma$-Ce
requires its inclusion.

Crucially, lattice dynamics in cerium are not independent of the correlated electronic
state. The strong renormalisation of the $4f$ electrons modifies the interatomic force
constants, so that phonon frequencies and vibrational entropies depend on whether the
material is in the $\alpha$ or $\gamma$ electronic state.
Conventional density-functional treatments of lattice dynamics, which do not incorporate the feedback of $4f$ correlations on the interatomic forces, cannot capture this effect.
Computational methods that combine DFT+DMFT with linear-response calculations of forces
have been developed for correlated materials.
{The calculation of phonons in correlated materials 
within DFT+DMFT was pioneered in the context of 
paramagnetic iron, where the frozen-phonon method was 
employed to demonstrate that electronic correlations 
have a strong effect on the lattice stability and 
phonon spectra across phase transitions~\cite{leonov2012calculated,leonov2014electronic}. 
The observed phonon softening in iron was subsequently 
attributed to the melting of ferromagnetic order rather 
than the structural transition itself, as revealed by 
eDMFT calculations~\cite{han2018phonon}. A framework for 
computing forces within DFT+DMFT that explicitly 
incorporates electronic entropy was established 
in Ref.~\cite{haule2016forces}, and an efficient scheme 
combining fixed self-energy calculations with 
non-diagonal supercells was developed and benchmarked 
on Fe, NiO, MnO, and SrVO$_3$~\cite{kocer2020}. 
Correlation-driven phonon anomalies have further been 
reported in FeSe~\cite{khanal2020correlation} and 
vanadium~\cite{pandey2023investigating}. For cerium specifically, 
DMFT-renormalised forces and phonon spectra were 
computed in Ref.~\cite{plekhanov2018many}. In all these 
cases, however, the focus was placed on the phonon 
spectrum itself, without the resulting phonon free 
energy being incorporated into the equation of state. 
In the present work, this gap is closed by 
systematically evaluating the DMFT-renormalised phonon 
free energy across the full volume range of the 
$\alpha$--$\gamma$ transition and demonstrating its 
quantitative role in determining the equilibrium 
lattice constant of $\gamma$-Ce.}

In this work, we close this gap by combining DFT+DMFT with a treatment of phonons
that incorporates the influence of correlated electrons.
We focus specifically on the vibrational free energy of $\gamma$-cerium
as renormalised by electronic correlations, and on its contribution to the equilibrium
lattice constant.
By treating electronic and vibrational degrees of freedom consistently within the
correlated framework, we extend the thermodynamic description of cerium to include the full free energy at finite temperature.

\section{Methods}
We carried out DFT combined with DMFT calculations at an inverse temperature of $\beta = 20$ eV$^{-1}$, corresponding to $T \simeq 580$ K, employing the embedded DMFT functional package \textsc{eDMFT}~\cite{haule2015free} interfaced with the all-electron \textsc{WIEN2k} code~\cite{blaha2020wien2k}. This framework enables a fully self-consistent treatment of both the electronic structure and the lattice dynamics in elemental cerium within the same theoretical formalism.

The DMFT impurity problem was solved using a continuous time quantum Monte Carlo (CT-QMC) method~\cite{haule2007quantum,Gull2011continuous}. {The statistical errors of the CT-QMC calculations, estimated from the 
fluctuations of the total energy over the final self-consistent DMFT 
iterations, are of order $1$~meV per primitive cell, negligible on the 
energy scales discussed in this work.} Following Ref.~\cite{haule2015exact}, which shows that the $4f$ occupancy of elemental cerium is robust with respect to the choice of double counting scheme and remains close to its nominal value, we adopt the nominal double counting correction\cite{haule2010dynamical}
%
\begin{equation}
V_{\mathrm{dc}} = U\left(n_f - \tfrac{1}{2}\right) - J\left(\tfrac{n_f}{2} - \tfrac{1}{2}\right)
\end{equation}
with fixed $n_f = 1$ throughout this work. 
The resulting $4f$ occupancy obtained from the fully converged DFT+DMFT calculations is discussed in Section S1 of Supplementary Information (SI), where it is shown to remain close to unity over the entire range of lattice constants considered. For further technical details of the implementation, we refer to Ref. ~\cite{haule2015free}. The Matsubara self-energy was analytically continued to the real frequency axis using the maximum entropy method~\cite{kaufmann2023ana_cont}. The quasiparticle weight was independently extracted from both Matsubara and real frequency data and found to be mutually consistent, providing an internal validation of the calculations.

In the \textsc{WIEN2k} calculations, the basis-set size was controlled by choosing RKmax = 7, which is commonly adopted for cerium and provides a well-converged description of total energies within the present setup \cite{huang2019electronic}. 
DFT calculations were carried out within both the local density approximation (LDA) \cite{perdew1992accurate} and the Perdew–Burke–Ernzerhof generalised gradient approximation (PBE) \cite{perdew1996generalized}.
Within DFT + DMFT, the cerium $f$ orbitals were described using an on-site Coulomb interaction of $U = 6.0$ eV and a Hund coupling of $J = 0.7$ eV, consistent with values successfully employed in previous studies of elemental cerium~\cite{held2001cerium,amadon2006alpha,haule2007quantum,huang2019electronic} as well as related Ce-based compounds~\cite{tomczak2018thermoelectricity}. The dependence of the equilibrium lattice constant on the choice of $U$ within DMFT is discussed in detail in Section S2 of SI.  The local Coulomb interaction, parametrised by $U$ and $J$, is treated within the DMFT impurity solver in the density-density (Ising) form \cite{haule2010dynamical}, consistent with Ref.~\cite{haule_tutorial3}. 

For the lattice dynamics calculations, a finite displacement approach is employed within an in-house framework that provides a flexible interface with both DFT and DFT + DMFT~\cite{kocer2020,chen2026impact}.
Since each displaced supercell in DFT+DMFT requires a full solution of the impurity problem and the computational cost increases with supercell size, the non-diagonal supercell technique \cite{lloyd2015} is adopted to sample a $4 \times 4 \times 4$ $\mathbf{q}$-point grid efficiently. 
Phonon calculations are performed at fixed lattice constants rather than at fully relaxed equilibrium volumes. In elemental cerium, this procedure does not generate spurious forces within the finite-displacement framework. Owing to the single-atom basis of the fcc structure and the preservation of crystal symmetry, a uniform change of the lattice constant produces only a finite hydrostatic stress, while the internal atomic forces remain strictly zero. {The equilibrium volume is therefore determined directly from the equation of state through a systematic scan of lattice constants.} The harmonic phonon problem is well defined, and no imaginary modes are observed across the volume range considered.
Throughout this work, the fixed self-energy approximation \cite{kocer2020} is employed for all displaced configurations. Within this approach, the DMFT self-energy is computed once for the equilibrium structure and subsequently kept fixed when evaluating the forces for the displaced geometries, thereby neglecting the explicit displacement dependence of the local self-energy. {This approximation reduces the computational 
cost by approximately one order of magnitude compared to a fully 
self-consistent variable self-energy scheme, as the latter 
requires a complete DMFT self-consistency cycle for each 
displaced supercell configuration.} The validity of this approximation is assessed in Section S3 of SI through a comparison with phonon calculations performed using a variable self-energy scheme.
In elemental cerium, spin--orbit coupling (SOC) governs 
the fine multiplet structure and effective degeneracy of 
the 4f states, which are central to the electronic 
free-energy landscape associated with the $\alpha$--$\gamma$ 
transition~\cite{lanata2013gamma,haule2015free}. For this 
reason, SOC is explicitly included in all calculations of 
the internal energy and electronic free energy presented 
in this work. {We note that its quantitative effect 
on the electronic free energy as a function of lattice 
constant remains small, as discussed in the RESULTS section.}
%
%
Spin-orbit coupling (SOC) \cite{koelling1977technique} was not included in the DMFT phonon calculations, as the current implementation lacks the interface to treat SOC within the finite-displacement phonon workflow. A similar approximation was adopted in Ref.~\cite{chen2026impact}.
To quantify the impact of this approximation, we examine the effect of SOC on the DFT+DMFT spectral function, which is further discussed in Section S4 of SI.
While SOC substantially reshapes the low-energy electronic spectrum in the vicinity of the Fermi level, the corresponding correction to the phonon free energy is small on the relevant thermodynamic energy scale on the level of DFT, as demonstrated in Section S5 of SI. { On the DFT+DMFT level, SOC slightly expands the equilibirum lattice constant, but restoring forces (and thus phonon frequencies) are largely unaffected, as shown in Section S6 of SI.} In all, these results indicate that, although SOC is essential for an accurate description of the electronic structure, its influence on the lattice-dynamical contribution to the total free energy remains secondary within the present framework.
In this work, the DMFT electronic free energy is evaluated using the stationary implementation of the DFT+DMFT functional \cite{haule2015free}. Crucially, this formalism goes beyond the internal energy $E$ by explicitly evaluating the Luttinger-Ward functional, which inherently incorporates the electronic entropy $S$ of the strongly correlated 4f shell, approximated by the impurity entropy obtained from the CT-QMC solver.
The phonon free energy $F_{ph}(T)$ is calculated within the harmonic approximation by summing over all phonon modes in the Brillouin zone. For a given temperature $T$, the absolute phonon free energy is expressed as:
\begin{equation}
F_{\mathrm{ph}}(T) = \frac{1}{2} \sum_{\mathbf{q}, \nu} \hbar \omega_{\mathbf{q}, \nu} 
+ k_{\mathrm{B}} T \sum_{\mathbf{q}, \nu} \ln \left[ 1 - \exp \left( -\frac{\hbar \omega_{\mathbf{q}, \nu}}{k_{\mathrm{B}} T} \right) \right]
\end{equation}
where $\omega_{\mathbf{q}, \nu}$ denotes the DMFT-renormalized phonon frequency for wave vector $\mathbf{q}$ and branch index $\nu$. The first term represents the zero-point energy contribution, while the second term captures the thermal effect through Bose-Einstein statistics \cite{wallace1972thermodynamics}. 

\section{Results}
\begin{figure}[t]
\includegraphics[width=1\columnwidth]{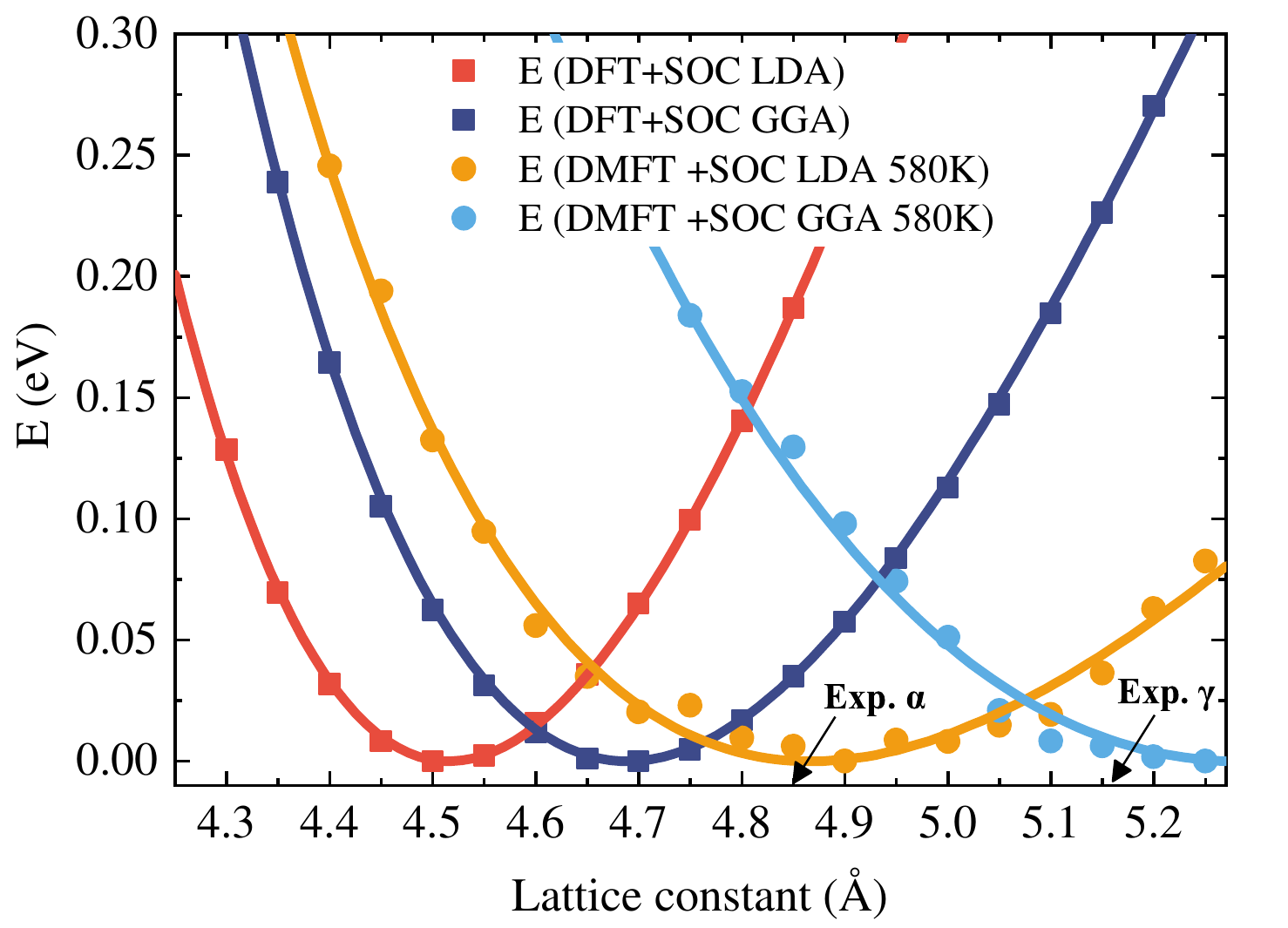}
\caption{Lattice-parameter dependence of the internal energy of cerium, calculated within DFT and DMFT.
Experimental lattice constants at ambient pressure are indicated by arrows: Exp. $\alpha$ = 4.85 Å (low temperature) \cite{koskenmaki1978cerium} and Exp. $\gamma$ = 5.16 Å (room temperature) \cite{olsen1985crystal}.
}
\label{dft-dmft-lattice-vs-energy}
\end{figure}

\begin{figure*}[t]
\includegraphics[width=1\textwidth]{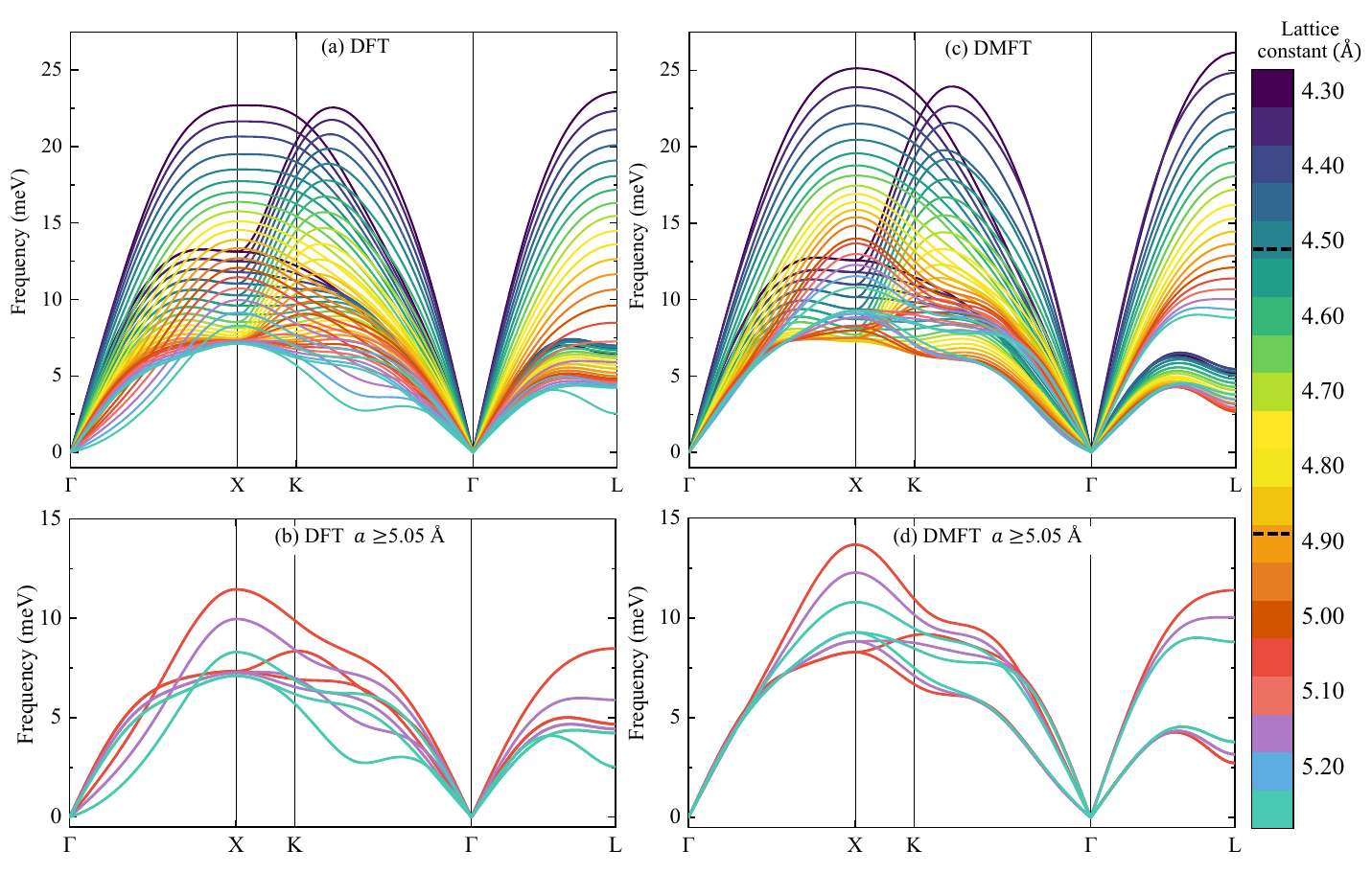}
\caption{ Ab initio computation of correlation-driven phonon dispersions in cerium. (a, c) Full phonon dispersion relations explicitly calculated from first principles within standard DFT (LDA, $T = 0$) and DFT+DMFT ($T = 580$ K), using a $4 \times 4 \times 4$ $\mathbf{q}$-point mesh. The colour bar indicates the lattice constant (in Å), with dashed lines marking the respective theoretical equilibrium constants ($a_0^{\text{DFT}} = 4.52$ Å; $a_0^{\text{DMFT}} = 4.88$ Å) from Fig.\ref{dft-dmft-lattice-vs-energy}.(b, d) Detailed views of the directly computed large-volume $\gamma$-phase regime ($a = 5.05$, $5.15$, and $5.25$ Å)}
\label{lda_ldadmftphonon}
\end{figure*}

\subsection{Electronic internal and electronic free energy}
The lattice-parameter dependence of the internal energy obtained within DFT and DFT+DMFT is shown in Fig. \ref{dft-dmft-lattice-vs-energy}. The calculated curves span a broad range of lattice constants covering both the compressed-volume regime associated with the $\alpha$-phase and the expanded-volume regime corresponding to the 
$\gamma$-phase. 

%


Within conventional DFT, including SOC, the equilibrium lattice constant at zero temperature is
\(a_0 = 4.52~\text{\AA}\) at the LDA+SOC level, in close agreement with Ref.~\cite{plekhanov2018many, casadei2012density}, where SOC was not included. 
The corresponding GGA calculation yields a larger equilibrium lattice constant of $a_0 = 4.69~\text{\AA}$, consistent with previous GGA results \cite{casadei2016density, casadei2012density}. As expected, the GGA functional predicts a larger equilibrium volume than LDA due to its reduced tendency to overbind \cite{mattsson2008am05}. 
From a fundamental perspective, density functional theory is a zero-temperature formalism and is therefore targeting the low-temperature $\alpha$ phase. Yet, both LDA and GGA severely underestimate its equilibrium lattice constant (4.52~Å and 4.69~Å, respectively, versus the experimental value of 4.85~Å \cite{koskenmaki1978cerium}). This pronounced discrepancy reveals the inability of static mean-field approaches to capture the strongly correlated nature of the 4f electrons, leading to an overestimation of binding. Consequently, there is no rigorous physical basis for expecting standard DFT to reproduce the significantly larger lattice constant of the entropy-stabilised high-temperature $\gamma$ phase (experimental $a \approx$ 5.16 Å \cite{olsen1985crystal}). These limitations motivate the incorporation of dynamical many-body correlations together with finite-temperature thermodynamic contributions to obtain a meaningful equation of state.

In addition, as shown in Section S6 of SI, the total-energy curves as a function of lattice constant obtained from LDA and LDA+SOC calculations are essentially identical. This indicates that SOC has a negligible effect on the equilibrium structural properties at the DFT level. 
Furthermore, the inclusion of SOC induces only minor modifications to the DMFT electronic free energy, as shown in Section S7 of SI, where the SOC and non-SOC free-energy curves nearly coincide over the entire lattice-parameter range at 580 K. 
%
Collectively, these findings demonstrate that SOC exerts a negligible influence on both the equilibrium lattice constants and the finite-temperature free-energy landscape within the present theoretical framework.
%

%

The inclusion of dynamical electronic correlations within DMFT leads to a pronounced lattice expansion. At \(T = 580\)~K, the LDA+DMFT+SOC minimum shifts to 
\(a_0 = 4.88~\text{\AA}\), substantially larger than the LDA+SOC value. The resulting lattice constant lies significantly closer to the experimental $\gamma$-phase value and reflects the reduction of the DFT overbinding tendency.
The overall behavior of the total energy curves is in close agreement with previous LDA+DMFT studies of cerium~\cite{plekhanov2018many,pourovskii2007self,haule2015free}, which reported comparable correlation-induced shifts of the equilibrium lattice constant at the level of the internal energy \(E\). In those works, the local \(4f\) correlations were treated using both the Hubbard I approximation and the CT-QMC impurity solver.  The quantitative consistency of the internal energy minima obtained here with earlier results demonstrates the robustness of the LDA+DMFT description of cerium.

In contrast, the GGA+DMFT+SOC calculation yields a substantially larger equilibrium lattice constant of $a_0 = 5.28$~\AA, which exceeds the experimental value, as noted previously\cite{haule2015free}. 
Indeed, the largest uncertainty in DFT+DMFT predictions of lattice constants comes from the choice of the DFT exchange-correlation potential (for other $f$-electron materials, see \cite{amadon2016first}).
{ Still, the temperature and correlation-induced lattice expansion within DFT+DMFT will also depend on the details of the many-body setting employed (interaction parameters, double-counting scheme and valency, temperature).
 The dependence of the GGA+DMFT+SOC equilibrium lattice constant on the choice of $U$ is examined in Section S8 of the SI, where we show that a reduced $U$ can bring the internal-energy lattice constant into agreement with experiment. 
 Alternatively, in an LDA+DMFT+SOC setting, a larger $U$ will improve the volume. The semi-empirical choice of $U$ is however constrained by congruence of spectral properties, for which previous DMFT studies indeed found $U = 6$~eV to be accurate.
}
The large variability for varying DFT potentials and Hubbard interactions significantly limits the predictiveness of the DFT+DMFT approach and calls for truly first principles approaches, such as $GW$+DMFT \cite{Biermann2003,Tomczak2017review}.
 
Here, instead, we will take a physics-informed decision. Below, we will be taking into account phonon contributions to the free energy. Phonon entropy generally favours larger equilibrium volumes~\cite{wallace1972thermodynamics}. Indeed, conversely, a larger lattice constant leads to softer interatomic force constants. The resulting lower phonon frequencies mean that these modes become more thermally accessible, yielding higher vibrational entropy.
 Since any DFT functional must yield an electronic equilibrium volume smaller than experiment, so that phonon entropy can account for the remaining expansion, we will proceed with LDA+DMFT, not GGA+DMFT.

%

\subsection{Lattice dynamics and phonon free energy}
We now extend the analysis by incorporating vibrational free-energy contributions into the energy landscape. 
%
Figure \ref{lda_ldadmftphonon}(a) represents the phonon dispersion relations calculated within the local density approximation (LDA) along the high-symmetry path
$\Gamma$--X--K--$\Gamma$--L for a series of lattice constants ranging from 4.30\AA\ to 5.25\AA.

For all lattice constants considered, the three acoustic phonon branches originate from zero frequency at the $\Gamma$ point, consistent with translational invariance.
No imaginary phonon frequencies are observed anywhere in the Brillouin zone for any of the lattice constants investigated.
This establishes that the crystal structure remains dynamically stable within the LDA description over the entire volume range examined, with no indication of soft-mode-driven lattice instabilities.

A clear and systematic volume dependence of the phonon spectrum is observed.
As the lattice constant increases, phonon modes exhibit a pronounced softening across the Brillouin zone.
At smaller lattice constants, corresponding to compressed volumes, the phonon frequencies are uniformly higher, reflecting stronger interatomic force constants.
Upon lattice expansion, the frequencies decrease smoothly and continuously, indicating a conventional weakening of bonding interactions rather than the emergence of anomalous lattice behaviour.

The softening is most pronounced for phonon branches in the intermediate- and high-frequency range, which show substantial downward shifts with increasing lattice parameter.
Despite this softening, all modes remain well separated from zero frequency, confirming the absence of incipient dynamical instabilities even at the largest lattice constants considered.
{The continuous evolution of these modes further confirms that the lattice response to volume expansion within the DFT description is governed by changes in the harmonic interatomic force constants, with no evidence of anharmonic or soft-mode-driven instabilities.}

Figure \ref{lda_ldadmftphonon}(b) presents a detailed view of the phonon dispersion relations of cerium in the large-volume $\gamma$-phase regime ($a \geq 5.05$ \AA), calculated within standard DFT (LDA; $T = 0$ K) along the high-symmetry path $\Gamma$--X--K--$\Gamma$--L. 
As the lattice constant increases from 5.05 \AA\ to 5.25 \AA, as indicated by the colour evolution from purple/red to cyan/green, a pronounced and highly non-uniform softening emerges in the low-frequency phonon branches. This effect is particularly evident in the low-energy modes and is most severe along the K--$\Gamma$ segment, where the branches exhibit a clear downward bending and a substantial reduction in frequency. 
The softening is markedly stronger in the low-frequency sector than in the high-frequency branches, which instead display a more gradual and nearly rigid shift. This behaviour, therefore, cannot be interpreted as a simple volume-induced rescaling of the phonon spectrum. Rather, it reflects a mode-selective renormalisation, with the low-energy vibrational modes showing enhanced sensitivity to lattice expansion. 
Such anomalous softening indicates that specific phonon modes, especially along X--K--$\Gamma$, are strongly affected by changes in the underlying bonding environment in the expanded lattice. This selective response highlights the nontrivial lattice-dynamical behaviour in the $\gamma$-phase, where the low-frequency modes undergo a significantly stronger renormalisation than the rest of the spectrum.

The absence of pronounced mode flattening or imaginary modes at the Brillouin-zone boundaries further excludes the presence of competing low-energy structural distortions within the explored parameter range.
Instead, the phonon spectrum retains well-defined dispersions and finite mode separations throughout, underscoring the robustness of the lattice against symmetry-lowering distortions driven by harmonic phonon instabilities. While higher-order anharmonic effects \cite{fultz2010vibrational} could quantitatively modify the high-temperature lattice dynamics, the complete absence of harmonic soft modes confirms the intrinsic mechanical stability of the fcc phase across the explored volume range.

 Figure \ref{lda_ldadmftphonon}(c) presents the phonon dispersions calculated within the LDA+DMFT framework along the same high-symmetry path and for the same lattice constants as those shown in (a) panel, enabling a direct,
mode-resolved comparison between the uncorrelated and correlated descriptions.
Again, the three acoustic branches emanate from zero frequency at $\Gamma$, and all phonon
frequencies remain positive throughout the Brillouin zone across the entire lattice-parameter range.
The absence of imaginary modes in both approaches indicates that the structure is dynamically stable
within the harmonic approximation and that the inclusion of local electronic correlations does not
introduce symmetry-lowering lattice instabilities. Quite the opposite, a phonon-minimum developing in DFT between $K$ and $\Gamma$ for the largest lattice constant, disappears in DMFT.

While the overall topology of the spectrum and its smooth volume evolution are preserved, LDA+DMFT
induces a clear quantitative renormalisation that is not uniform across branches. Most notably, the
highest branches are systematically shifted upward in LDA+DMFT relative to LDA, reaching
significantly larger peak frequencies, particularly near the zone-boundary region around X and along
the approach to L. This correlation-induced hardening of the upper phonon manifold is clearly evident in panel (c) of Fig.~\ref{lda_ldadmftphonon} relative to panel (a), and this effect persists over the entire range of lattice constants investigated. At the same time, the lower-lying branches
exhibit a more moderate and mode-dependent response. These branches show
comparatively smaller shifts with respect to LDA and, in several $q$-space regions, a tendency toward reduced dispersion
and enhanced clustering, indicating that correlations modify the relative separations of nearby modes
rather than producing a simple rigid shift of the entire spectrum.

The volume dependence remains conventional in both methods: increasing the lattice constant leads
to an overall softening trend.
However, the magnitude and distribution of this softening differ once correlations are included.
Within LDA+DMFT, changes in curvature and mode spacing are especially apparent along the X--K--$\Gamma$ segment, where several branches become more closely packed, and their dispersions evolve differently from the LDA case. Along the $\Gamma$--L direction, the comparison highlights a particularly strong
upward renormalisation of the highest-frequency branches in LDA+DMFT, while the lower branches remain closer to their LDA counterparts.

\begin{figure*}[t]
\includegraphics[width=1\textwidth]{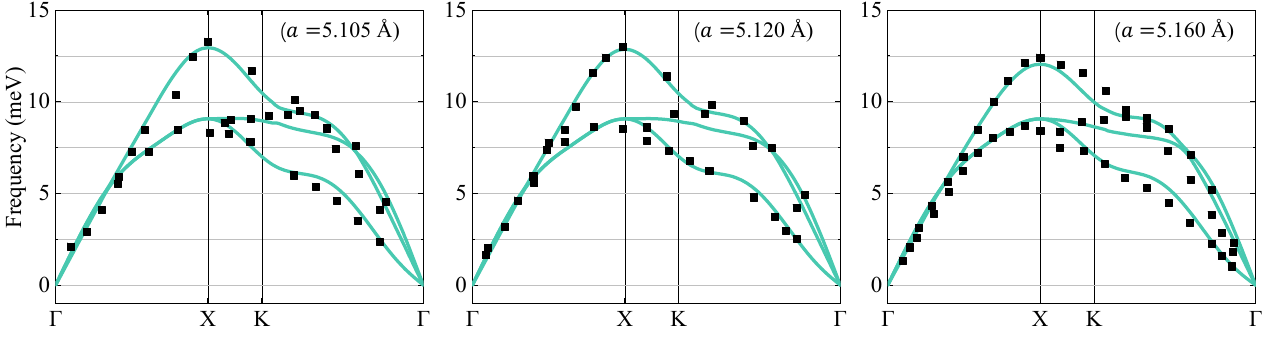}
\caption{Phonon dispersion relations obtained using principal component-based neural network interpolation trained on LDA+DMFT calculations, compared with experimental data from cerium under hydrostatic pressure at ambient temperature from Ref. \cite{krisch2011phonons}. Results are shown for lattice constants $a$=5.105, 5.120 and 5.160\AA\ along the $\Gamma$–X–K–$\Gamma$ high-symmetry path. The predicted dispersions (green lines) show very good agreement with the experimental phonon frequencies (black symbols). 
}
\label{ml_predict_3lattice}
\end{figure*}
Panel (d) of Figure \ref{lda_ldadmftphonon} presents a detailed, volume-resolved view of the phonon dispersion relations for cerium in the large-volume $  \gamma  $-phase regime ($  a = 5.05  $, $5.15$, and $5.25$ Å), extracted directly from the LDA+DMFT calculations. We focus specifically on this expanded lattice-parameter range because it corresponds to the physical state of the high-temperature $  \gamma  $-phase, where the influence of electronic correlations on lattice dynamics is most dramatic and physically relevant.
A central finding from our LDA+DMFT results is that the individual vibrational branches respond in a highly non-uniform and mode-selective manner to lattice expansion — a complex behaviour that is completely absent in the standard DFT description. 
%
In typical DFT calculations (see panel (b) of Fig.~\ref{lda_ldadmftphonon}), phonon branches undergo a conventional volume-driven softening, with the low-frequency branches exhibiting a particularly anomalous and rapid decrease in frequency. In striking contrast, the LDA+DMFT dispersions in panel (d) reveal a remarkable inversion precisely in the low-frequency region. While the high-frequency branches still follow the expected trend and soften with increasing lattice parameter, the low-frequency branches at the X and L points actually move upwards. Most notably, at the X point, the frequency rises from 8.28\,meV to 9.28\,meV as the lattice constant expands from 5.00\,\AA{} to 5.25\,\AA{}---an increase of more than 1\,meV over this volume range. {This upward shift is carried specifically by the transverse acoustic (TA) branch at the X point, while the longitudinal acoustic (LA) branch continues to soften conventionally, decreasing from approximately 13.7~meV to 10.8~meV over the same volume range. The TA branch stiffens while the LA branch softens, and this mode-selective response reflects a correlation-driven renormalisation of the transverse interatomic force constants by the localised 4f electrons. This renormalisation counteracts the usual volume-induced softening specifically for shear-type atomic displacements in the $\gamma$-phase. The anisotropic character of this effect confirms that 4f correlations modify the interatomic force constants in a way that cannot be captured by conventional DFT.} 

{The physical origin of this correlation-induced hardening can be understood in terms of the renormalisation of electronic screening by the localised 4f shell. Within DFT, the partially itinerant 4f electrons contribute to the screening of interatomic forces under transverse atomic displacements, effectively reducing the restoring force and softening the transverse force constants. The DMFT self-energy suppresses this itinerant contribution by dynamically transferring 4f spectral weight away from the Fermi level (see Section S4 of SI for the corresponding spectral function), thereby reducing the capacity of the 4f electrons to screen shear-type displacements. As a consequence, the transverse interatomic force constants are stiffened relative to their DFT counterparts, producing the observed hardening of the TA branches in the $\gamma$-phase volume regime. The methodological framework underlying such DMFT-renormalised forces has been developed in Refs. \cite{leonov2014first, haule2016forces, plekhanov2018many}, and applied to a range of correlated systems including paramagnetic iron and cerium compounds; the mode-selective TA hardening identified here represents a specific manifestation of this many-body renormalisation in the $\gamma$-phase of elemental Ce.}

The comparison between panels (b) and (d) of Fig.~\ref{lda_ldadmftphonon} is therefore particularly illuminating: the uncorrelated DFT picture shows an anomalously rapid downward plunge of the low-energy modes, whereas the inclusion of dynamical correlations in DMFT not only suppresses that softening but actively reverses it into an upward shift for the very same branches. This inversion confirms that the low-frequency vibrations in these regions of the Brillouin zone are extremely sensitive to correlation-driven modifications of the local bonding geometry. In the strongly correlated $  \gamma  $-phase, the 4$  f  $ electrons renormalise the force constants in a mode-specific way, leading to a non-trivial redistribution of phonon energies that cannot be captured by conventional DFT.
Such correlation-induced lattice-dynamical effects are crucial for the thermodynamic stability of the expanded $\gamma$-phase and highlight the necessity of treating electronic correlations and phonon renormalisation on equal footing.

To quantify global versus local changes in phonon renormalisations, we
map the phonon dispersion calculated at a given lattice constant onto a high-dimensional vector of phonon frequencies along the chosen high-symmetry path. A principal component analysis then reveals that the dominant fraction of the variance is captured by only a few components, with the leading mode accounting for approximately 98\% of the total variance and corresponding to an almost uniform hardening or softening of the spectrum. 
Therefore, the correlation-induced modification of the phonon spectrum exhibits a pronounced collective character, implying that its lattice-parameter dependence can be represented on a low-dimensional manifold. 

Exploiting this reduced representation, we construct a continuous interpolation of the phonon dispersions as a function of lattice constant by modelling the projection coefficients of the leading principal components with a multilayer perceptron neural network. The interpolation is validated through leave-one-out tests, demonstrating excellent agreement with explicitly calculated phonon spectra and confirming that the smooth, collective evolution of the dispersions is faithfully captured. This compact description enables a numerically stable evaluation of the phonon free energy over the full lattice-parameter range considered. Further technical details of the interpolation procedure are provided in Section S9 of SI.

To further validate the interpolation scheme, DMFT phonon dispersions were predicted at three lattice constants that were not included in the training dataset and compared with experimental measurements reported in Ref. \cite{krisch2011phonons}. As shown in Fig.~\ref{ml_predict_3lattice}, the spectra predicted by the model trained on LDA+DMFT calculations exhibit excellent agreement with the experimental phonon branches along the high symmetry path. This consistency demonstrates that the low-dimensional representation, combined with neural network interpolation, reliably captures the collective evolution of the phonon spectrum and provides predictive capability beyond the explicitly computed data.

\begin{figure}[t]
\includegraphics[width=1\columnwidth]{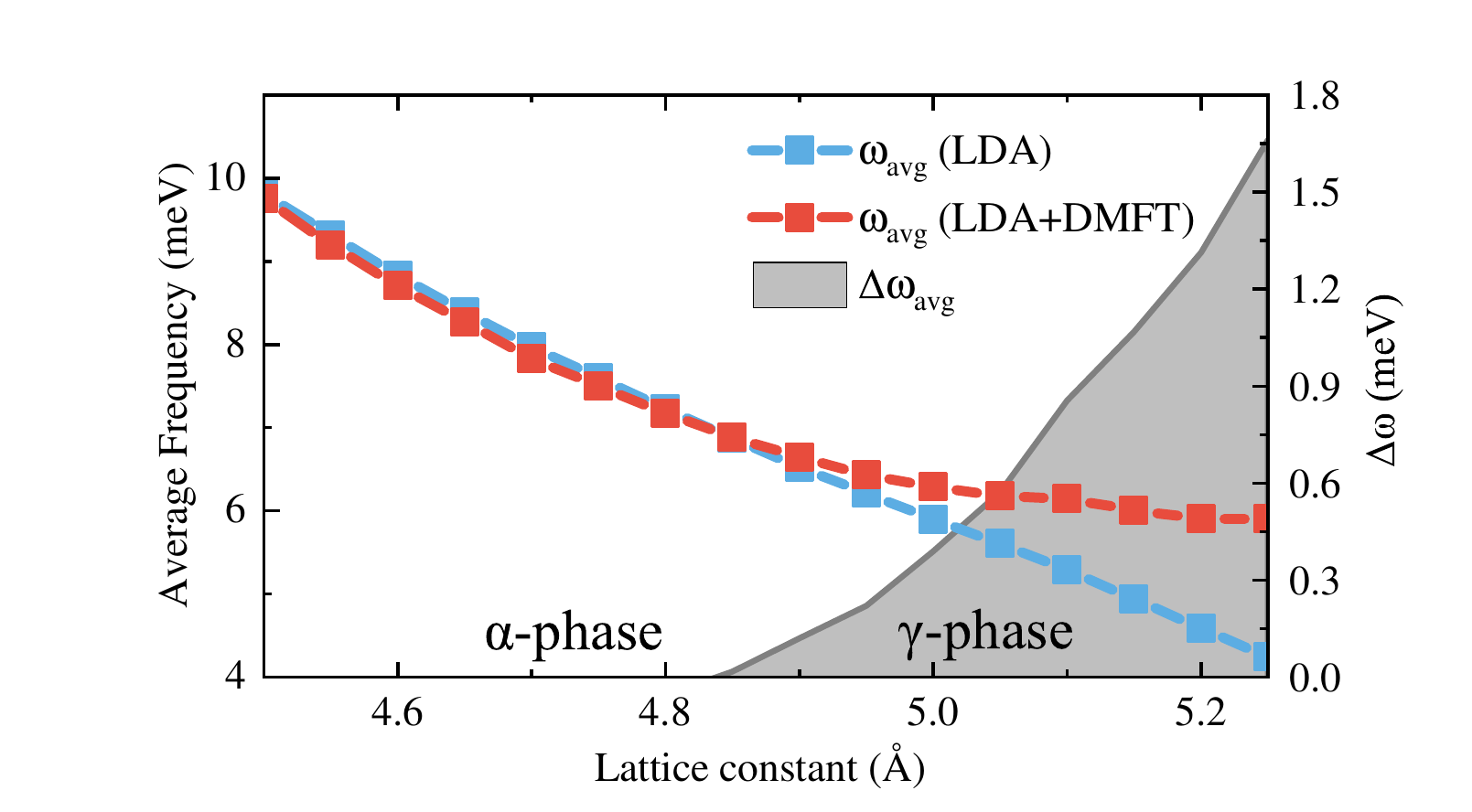}
\caption{Average phonon frequency as a function of lattice constant for elemental cerium, calculated within DFT (blue squares) and DFT+DMFT (red squares). The shaded region represents the difference in the average phonon frequency between the two approaches, defined as $\Delta \omega = \omega_{\mathrm{avg}}^{\mathrm{LDA+DMFT}} - \omega_{\mathrm{avg}}^{\mathrm{LDA}}$.
In the compressed volume regime corresponding to the $\alpha$-phase, the two methods yield nearly identical average frequencies, indicating a minor influence of electronic correlations on lattice dynamics. Upon lattice expansion towards the $\gamma$-phase, a systematic deviation develops, with LDA+DMFT suppressing the volume-induced phonon softening observed at the LDA level.}
\label{average_frequency_difference}
\end{figure}

 Although a phonon spectrum formally comprises a large number of frequencies associated with different wave vectors and vibrational branches, the above principal components analysis suggests that variations in volume primarily lead to shifts of groups of phonon branches, rather than independent changes of individual modes.
 It is then instructive to examine the lattice-parameter dependence of the overall phonon dispersion. 
 
Figure~\ref{average_frequency_difference} demonstrates a pronounced correlation-dependent modification of the lattice dynamics across the $\alpha$--$\gamma$ transition of the {\it average} phonon frequency. The average is obtained from the full phonon dispersions calculated within LDA and LDA+DMFT, shown in Fig.~\ref{lda_ldadmftphonon} ((a) and (c) panels, respectively), by averaging the phonon frequencies along the chosen high symmetry path in the Brillouin zone.
In the compressed volume regime corresponding to the $\alpha$-phase, the two approaches yield nearly identical average frequencies, such that $\Delta \omega_{\mathrm{avg}}$ remains close to zero. This indicates that electronic correlations exert only a marginal influence on the lattice response at small volumes.

With increasing lattice parameter, a systematic deviation emerges. Whereas LDA predicts substantial, near-linear phonon softening upon lattice expansion, the inclusion of dynamical electronic correlations within LDA+DMFT markedly reduces this softening, leading to a monotonically increasing frequency difference $\Delta \omega_{\mathrm{avg}}$. This behaviour demonstrates that electronic correlations qualitatively modify the lattice dynamics across the $\alpha$–$\gamma$ transition. The resulting non-linear evolution of the phonon frequencies with lattice constant further suggests that simple interpolation between discrete lattice constants may be unreliable, motivating the use of the machine-learning interpolation scheme introduced above.

\begin{figure}[t]
\includegraphics[width=0.88\columnwidth]{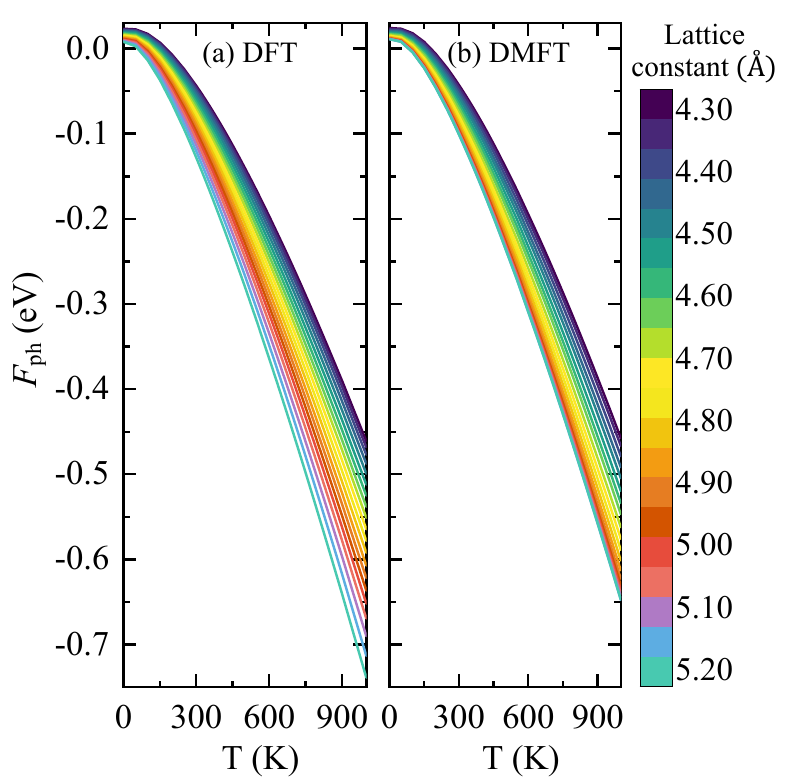}
\caption{{Temperature dependence of the absolute phonon free energy, $F_{\mathrm{ph}}$, of elemental cerium calculated within (a) LDA and (b) LDA+DMFT for lattice constants between 4.30 and 5.25~\AA. The LDA+DMFT phonon dispersions are computed at $\beta = 20$ ($T \approx 580$~K) and used within the harmonic approximation.}}
\label{cerium_lattice_free}
\end{figure}

Figure~\ref{cerium_lattice_free} shows the temperature dependence of the absolute phonon free energy,
$F_{\mathrm{ph}}(T)$, of elemental cerium calculated within the LDA framework (panel a) and LDA+DMFT framework (panel b) for lattice constants between 4.30 and 5.25~\AA.
The free-energy values have been obtained from explicit evaluation of the harmonic
free-energy expression at discrete temperature intervals of 50~K, with neighbouring points
connected, rather than from a global functional fit.
In the LDA+DMFT case, the underlying phonon dispersions are computed at a fixed inverse
temperature $\beta = 20$ (corresponding to $T \approx 580$~K), and are assumed to be
temperature-independent within the harmonic approximation.
The temperature dependence of $F_{\mathrm{ph}}(T)$ therefore arises solely from the Bose–Einstein
occupation factors entering the harmonic free-energy expression.
In both approaches, $F_{\mathrm{ph}}(T)$ decreases smoothly and monotonically with increasing
temperature from the low-temperature regime up to beyond 900~K.
This behavior reflects the progressive increase of the vibrational entropy contribution
and is consistent with a harmonically stable lattice over the entire lattice-parameter range.

In the LDA results shown in panel a of Fig. \ref{cerium_lattice_free}, the phonon free energy displays a pronounced
dependence on the lattice parameter. At a given temperature, curves corresponding to
larger lattice constants are systematically shifted to lower (more negative) values of
$F_{\mathrm{ph}}$, while smaller lattice constants yield higher free energies. This
ordering is preserved across the entire temperature range and becomes progressively more
pronounced at elevated temperatures. As temperature increases, the set of curves fans
out, indicating that the volume dependence of the vibrational free energy is strongly
enhanced by thermal effects. In particular, at temperatures above several hundred
kelvin, the separation between curves associated with neighbouring lattice parameters
becomes substantial, signalling a sizable contribution of phonons to the relative
thermodynamic stability of different volumes within the LDA description.

Panel b of Fig.~\ref{cerium_lattice_free} shows the corresponding phonon free
energies obtained within LDA+DMFT. As in the LDA case, all curves decrease smoothly with
temperature and remain well behaved throughout the entire temperature range, consistent
with the absence of dynamical instabilities as inferred from the phonon spectra. The overall
temperature dependence of $F_{\mathrm{ph}}(T)$ is qualitatively similar to that found in
LDA, indicating that the dominant entropic contribution associated with lattice
vibrations is preserved when local electronic correlations are included.

A clear quantitative difference emerges, however, in the volume dependence of the phonon
free energy. In contrast to the LDA results, the LDA+DMFT curves are significantly more
closely spaced for different lattice constants at all temperatures. Although the
monotonic ordering with lattice constant is maintained, the overall spread of
$F_{\mathrm{ph}}(T)$ is markedly reduced. This effect is particularly evident at high
temperatures, where the LDA results show a strong fan-out of curves, while the LDA+DMFT
curves remain comparatively clustered. The reduced separation between curves indicates
that the sensitivity of the phonon free energy to changes in lattice constant is
substantially suppressed once electronic correlations are taken into account.
This behavior is fully consistent with the correlation-induced
renormalisation of the phonon spectrum discussed earlier, which redistributes vibrational
frequencies in a nonuniform manner.

\begin{figure}[t]
\includegraphics[width=1\columnwidth]{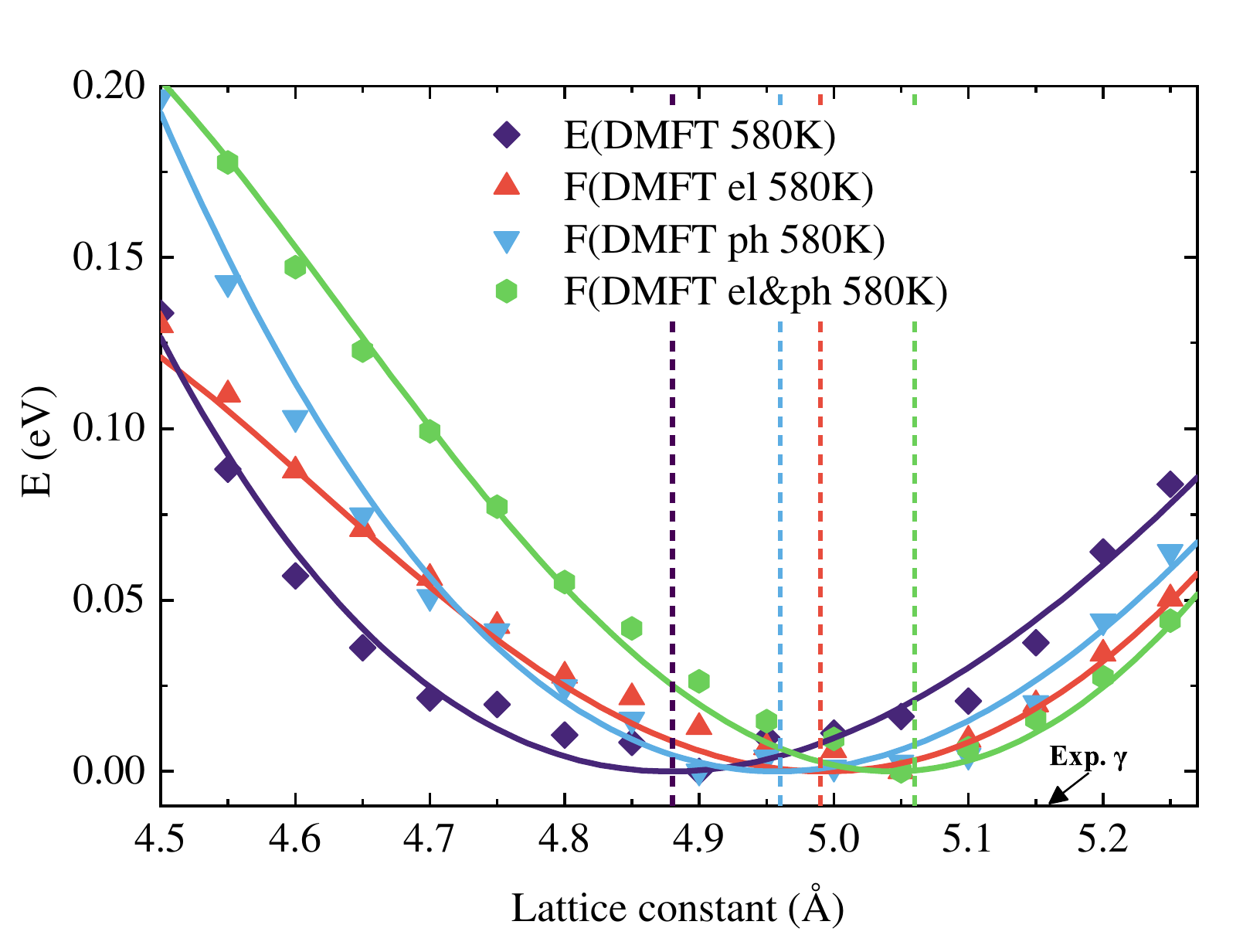}
\caption{
Lattice-parameter dependence of the internal energy and free-energy contributions within DMFT at 580~K. 
Symbols denote the calculated data points for the internal energy $E$, the electronic free energy $F_{\mathrm{el}}$, the phonon free energy $F_{\mathrm{ph}}$, and the combined electronic plus phononic free energy $F_{\mathrm{el+ph}}$. 
Solid lines represent third-order Birch--Murnaghan fits to the corresponding datasets. 
Vertical dashed lines indicate the lattice constants at which each fitted curve attains its minimum. All energies are given per primitive unit cell (one atom per fcc cell), and all curves have their minima aligned to zero.}

\label{dmft_580k}
\end{figure}

\begin{table*}[htbp]
\centering
\caption{Calculated and experimental fcc lattice constants of $\gamma$-Ce obtained using different theoretical approaches and experimental conditions.}
\label{tab:ce_lattice_comparison}
\begin{tabular}{lcccccl}
\hline\hline
Method & $U$ (eV) & $J$ (eV) & $T$ (K) & $P$ (GPa) & $a$ (\AA) & Ref. \\
\hline
\multicolumn{7}{c}{This work} \\
\hline
LDA+SOC &  &  &  &  & 4.52 & \\
GGA+SOC &  &  &  &  & 4.69 & \\
LDA+DMFT+SOC (internal $E$, CT-QMC) & 6.0 & 0.70 & 580 &  & 4.88 & \\
LDA+DMFT+SOC (electronic $F$, CT-QMC) & 6.0 & 0.70 & 580 &  & 4.99 &  \\
LDA+DMFT+SOC (phononic $F$, CT-QMC) & 6.0 & 0.70 & 580 &  & 4.96 &  \\
LDA+DMFT+SOC (electronic + phononic $F$, CT-QMC) & 6.0 & 0.70 & 580 &  & 5.06 &  \\
GGA+DMFT+SOC (internal $E$, CT-QMC) & 6.0 & 0.70 & 580 &  & 5.28 &  \\
\hline
\multicolumn{7}{c}{Prior theoretical work} \\
\hline
LDA &  &  &  &  & 4.49 & \cite{soderlind1994electronic} \\
LDA &  &  &  &  & 4.50 & \cite{plekhanov2018many} \\
LDA &  &  &  &  & 4.52 & \cite{amadon2008gamma} \\
LDA+DMFT non self-consistent ASA (Hubbard I) & 6.0 & 0.46 & 0 &  & 4.91 & \cite{pourovskii2007self} \\
LDA+DMFT self consistent ASA (Hubbard I) & 6.0 & 0.46 & 0 &  & 4.93 & \cite{pourovskii2007self} \\
LDA+DMFT (Hubbard I) & 6.0 & 0.70 & 232 &  & 4.95 & \cite{plekhanov2018many} \\
LDA+DMFT PAW (Hubbard I) & 6.0 & 0.00 & 273 &  & 4.98 & \cite{amadon2012self} \\
LDA+DMFT ASA (Hubbard I) & 6.0 & 0.00 & 273 &  & 4.91 & \cite{amadon2012self} \\
LDA+DMFT (internal $E$, Hirsch-Fye QMC) & 6.0 &     & 400 &  & 4.88 & \cite{amadon2006alpha}\\
LDA+DMFT (internal $E$, Hirsch-Fye QMC) & 6.0 &  & 800 &  & 4.99 & \cite{amadon2006alpha}\\
LDA+DMFT (internal $E$, Hirsch-Fye QMC) & 6.0 &  & 1600 &  & 5.03 & \cite{amadon2006alpha}\\
LDA+DMFT+SOC (internal $E$, CT-QMC) & 6.0 & 0.70 & 400 &  & 4.96 & \cite{haule2015free} \\
LDA+DMFT+SOC (internal $E$, CT-QMC) & 6.0 & 0.70 & 900 &  & 4.97 & \cite{haule2015free} \\
LDA+DMFT+SOC (electronic $F$, CT-QMC) & 6.0 & 0.70 & 400 &  & 5.04 & \cite{haule2015free} \\
LDA+DMFT+SOC (electronic $F$, CT-QMC) & 6.0 & 0.70 & 900 &  & 5.06 & \cite{haule2015free} \\
LDA+$U$ (FLL, FM imposed) & 6.1 & 0.70 &  &  & 5.04 & \cite{amadon2008gamma} \\
LDA+$U$ (AMF, FM imposed) & 6.1 & 0.70 &  &  & 4.99 & \cite{amadon2008gamma} \\
GGA &  &  &  &  & 4.70 & \cite{soderlind1994electronic} \\
GGA+$U$ (Dudarev, FM imposed) & 5.4 &  &  &  & 5.05 & \cite{amadon2008gamma} \\
GGA+$U$ (FLL, FM imposed) & 6.1 & 0.70 &  &  & 5.27 & \cite{amadon2008gamma} \\
r$^2$SCAN &  &  &  &  & 4.65 & \cite{giri2025isostructural} \\
PBE0 &  &  &  &  & 5.22 & \cite{casadei2016density} \\
SIC LSD &  &  &  &  & 5.14 & \cite{svane1996electronic} \\
LAK meta-GGA &  &  &  &  & 5.14 & \cite{giri2025isostructural} \\
\hline
\multicolumn{7}{c}{Experiment} \\
\hline
X ray diffraction ($\gamma$-phase) &  &  & room temperature & 0.00 & 5.16 & \cite{olsen1985crystal} \\
X ray diffraction ($\gamma$-phase) &  &  & room temperature & 0.56 & 5.10 & \cite{ma2016structure} \\
X ray diffraction ($\gamma$-phase) &  &  & room temperature & 1.30 & 5.06 & \cite{olsen1985crystal} \\
X ray diffraction ($\alpha$-phase)
&  &  & 77 & 0.00 & 4.85 & \cite{koskenmaki1978cerium} \\
X ray diffraction ($\alpha$-phase)
&  &  & 298 & 0.81 & 4.82 & \cite{koskenmaki1978cerium} \\
\hline\hline
\end{tabular}
\end{table*}

Since the DFT equilibrium volume deviates substantially from the experimental value, the inclusion of vibrational contributions at this level does not restore agreement nor modify the central conclusions of the present work. For clarity of presentation and to maintain focus on the correlation-driven effects captured within DFT+DMFT, the detailed temperature-dependent total free-energy analysis at the DFT level is therefore discussed in detail in Section S10 of SI.

Figure \ref{dmft_580k} illustrates the lattice-parameter dependence of the internal energy and the various free-energy contributions obtained within the DMFT framework at 580 K. In the absence of any entropic corrections, the internal-energy curve exhibits a minimum at a lattice constant of approximately 4.88 Å, which serves as a reference for assessing 
entropic effects. Upon including electronic entropy corrections, the equilibrium lattice constant grows to about 4.99 Å, indicating a moderate lattice expansion driven by local-moment entropy and highlighting the importance of finite-temperature electronic effects in the equation of state of cerium.

Instead, when only phonon free-energy contributions are incorporated (neglecting SOC), the minimum of the free-energy curve is located at a slightly smaller lattice constant of about 4.96 Å, showing that lattice vibrational effects alone do not stabilise the expanded structure either. By contrast, when both electronic and phononic contributions are taken into account simultaneously, the total free energy exhibits a minimum at approximately 5.06 Å, in good agreement with experiment.

This congruence underscores the crucial role of electronic correlations {\it and} vibrational entropy, both of which are captured within the current DMFT framework and are necessary to stabilise the high temperature, large volume $\gamma$-phase of cerium. We note that the DMFT free energy evaluated at the experimental lattice constant, $a=5.16$\AA\ \cite{olsen1985crystal}, exceeds the theoretical minimum by only about 0.015 eV per primitive cell $\approx 150 \mathrm{K} \times k_B$, which is small compared to the temperature, $T=580$~K, considered here. 

As shown in Table~\ref{tab:ce_lattice_comparison}, conventional LDA and GGA approaches significantly underestimate the lattice constant of $\gamma$-Ce. Earlier LDA+DMFT calculations based on the Hubbard-I approximation already yield improved values around 4.9--5.0~\AA. Given that these evaluations were effectively zero-temperature internal energy minimisations, it is physically expected that they fall slightly short of the experimental high-temperature volume.
%
Finite temperature LDA+DMFT calculations using QMC found slightly larger lattice constants on the basis of total internal energies, but still fell short of the experimental value.\cite{haule2015free}
More recent CT-QMC-based LDA+DMFT+SOC studies computing {\it electronic} free energies (using the impurity entropy), predict equilibrium lattice constants of 5.04~Å (5.06~Å) at $T=400$~K ($T=900$~K)~\cite{haule2015free}, in notably better agreement with experiment. In the present results, the lattice constant in a purely electronic picture comes out slightly smaller, $4.99$~\AA, at $T=580$~K. We believe this discrepancy to mainly arise from a different treatment of the Coulomb interaction. {While we use density-density type of interactions, Haule and Birol \cite{haule2015free} used the full Coulomb interaction.}
Combining the electronic free energies with the renormalised phonon free energy, we find the equilibrium lattice constant of $5.06$~\AA\ at $T=580$~K, in near-quantitative agreement with experiment. The phonon-driven $\sim 1.5$\% increase in lattice constant underscores the importance of a fully finite-temperature treatment that accounts for both electronic correlations and lattice vibrations. Future works will assess whether the remaining $1\%$ underestimation of the lattice constant can be accounted for when using the full Coulomb interaction along with SOC in our comprehensive electron-phonon workflow, or whether a step beyond density-functional based kinetics is required.

To assess the temperature robustness of the DMFT results, we have also examined the lattice-parameter dependence of the internal energy and the electronic free energy at 900 K. The corresponding data are presented in Section S11 of SI. The positions of the free-energy minima remain very close to those obtained at 580 K, with only a marginal shift towards larger lattice parameters, confirming that the conclusions drawn in the main text are insensitive to the specific temperature chosen within the $\gamma$-phase regime.

Also, it is worth noting that the agreement obtained with SIC LSD and the LAK meta-GGA \cite{svane1996electronic,giri2025isostructural} should be interpreted with caution, as neither approach includes an explicit treatment of dynamical correlations or finite-temperature entropy, both of which are essential for stabilising the $\gamma$-phase of cerium.

\section{Conclusion}
In this work, we have presented a comprehensive finite temperature investigation of the equilibrium lattice properties of elemental cerium by combining density functional theory with dynamical mean field theory and explicit phonon free energy contributions. By treating electronic correlations and lattice vibrations on an equal footing, we construct a correlated electron lattice free energy framework that enables a quantitative assessment of structural stability beyond static total energy considerations.

While conventional density functional approximations significantly underestimate the equilibrium volume of cerium, we find that the inclusion of dynamical electronic correlations within DMFT leads to a substantial expansion of the lattice. Importantly, electronic correlations also strongly renormalise the lattice dynamics, modifying both the phonon spectrum and its volume dependence. As a consequence, the phonon free energy exhibits a markedly reduced sensitivity to lattice constant once correlations are included, in sharp contrast to the pronounced volume dependence obtained at the DFT level.

The combined effect of electronic and vibrational entropy results in a pronounced restructuring and flattening of the finite temperature free energy surface in the expanded volume regime. This behaviour stabilises lattice constants that are in close quantitative agreement with experimental measurements of the high temperature $\gamma$-phase.


Specific to cerium, logical next steps are the inclusion of the full Coulomb interaction and spin-orbit coupling in all parts of the calculations, as well as simulations at lower temperatures to assess the emergence of a free-energy minimum associated with the $\alpha$-phase.

More broadly, our results demonstrate that correlation-induced renormalisation of phonon free energies constitutes a key ingredient for a physically controlled description of finite temperature phase stability in strongly correlated materials. The framework established here offers a systematic route for extending first-principles free energy calculations to complex correlated solids, where the interplay between electronic correlations and lattice dynamics is crucial.

\section*{Acknowledgments}
The authors would like to thank B. Monserrat, S. Poncé, L. Kantorovich and J. Kolorenč for useful discussions. Y.W. acknowledges support from the Marie Skłodowska Curie Actions COFUND programme under the Horizon Europe Programme through the Physics for Future (P4F) Programme (Grant Agreement No. 101081515). S.C. acknowledges support from EPSRC [EP/V062654/1], the Cambridge Trust and the Winton Program for the Physics of Sustainability. Computational resources were provided through our membership in the UK's HEC Materials Chemistry Consortium, funded by EPSRC (EP/R029431 and EP/X035859). This work utilised ARCHER2, the UK National Supercomputing Service (http://www.archer2.ac.uk), as well as resources from the UK Materials and Molecular Modelling Hub (MMM Hub), which is partially funded by EPSRC (EP/T022213/1,
    EP/W032260/1 and EP/P020194/1).  

\bibliography{apssamp}

\end{document}


\title{Supporting Information\\Correlation-driven phonon renormalisation and the equation of state of  $\gamma$-cerium}
\author{Yao Wei}
\email{yao.wei@kcl.ac.uk}
\affiliation{Theory and Simulation of Condensed Matter
(TSCM), King's College London, Strand, London WC2R 2LS, United Kingdom}
\affiliation{Institute of Physics (FZU), Czech Academy of Sciences, Na Slovance 2, 182 00 Praha, Czech Republic}
\affiliation{Institute of Informatics, Slovak Academy of Sciences, 845 07 Bratislava, Slovakia}
\author{Siyu Chen}
\affiliation{Cavendish Laboratory, University of Cambridge, Cambridge CB3 0HE,  United Kingdom}
\affiliation{Department of Materials Science and Metallurgy, University of Cambridge, Cambridge CB3 0FS,  United Kingdom}
\author{Evgeny Plekhanov}
\affiliation{Theory and Simulation of Condensed Matter
(TSCM), King's College London, Strand, London WC2R 2LS, United Kingdom}
\author{Ivan Štich}
\affiliation{Institute of Informatics, Slovak Academy of Sciences, 845 07 Bratislava, Slovakia}
\author{Cedric Weber}
\email{cedric.weber.phd@gmail.com}
\affiliation{Quantum Brilliance Pty, The Australian National University, Canberra, Australian Capital Territory 2600, Australia}
\author{Jan Tomczak}
\email{jan.tomczak@kcl.ac.uk}
\affiliation{Theory and Simulation of Condensed Matter
(TSCM), King's College London, Strand, London WC2R 2LS, United Kingdom}
\affiliation{Institute of Solid State Physics, TU Wien, Vienna, Austria}

\maketitle
\section{S1. Ce $4f$ occupancy as a function of lattice parameter}
\begin{figure}[h]
  \centering
  \includegraphics[width=0.7\columnwidth]{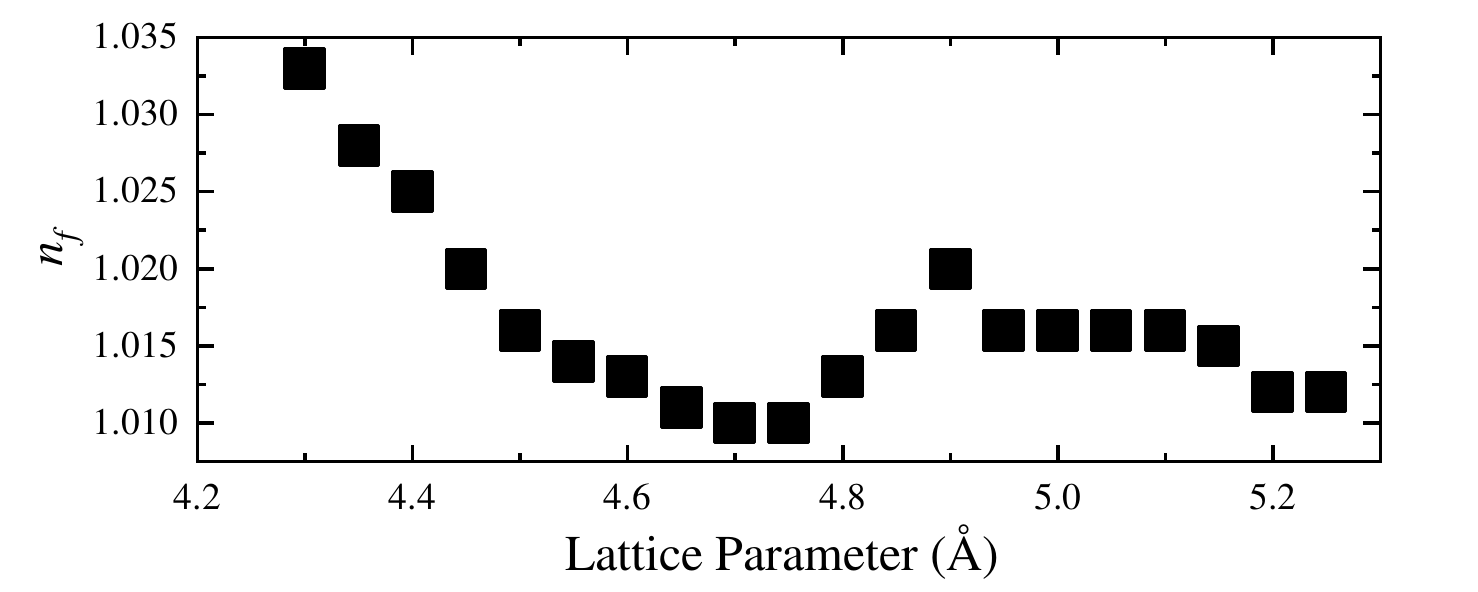}
  \caption{Ce $4f$ impurity occupancy $n_{f}$ obtained from the converged LDA+DMFT+SOC calculations as a function of the lattice parameter. Across the entire range of lattice parameters investigated, $n_{f}$ remains stable, staying within 3\% of single occupancy. }
  \label{nf}
\end{figure}

Figure~\ref{nf} shows the Ce $4f$ impurity occupancy $n_f$ obtained from the fully converged DFT+DMFT calculations as a function of the lattice parameter. Across the entire range of lattice parameters investigated, $n_f$ remains close to unity and exhibits only a weak dependence on the lattice parameter. No abrupt change or systematic drift of the $4f$ occupancy is observed. This demonstrates that the electronic configuration remains stable throughout the investigated volume range and that the phonon and free energy results discussed in the main text are not affected by spurious changes in the $4f$ occupancy.

\section{S2. Effect of Hubbard U on the Equilibrium Lattice Parameter within DMFT}
\begin{figure}[ht]
\centering
\includegraphics[width=0.49\columnwidth]{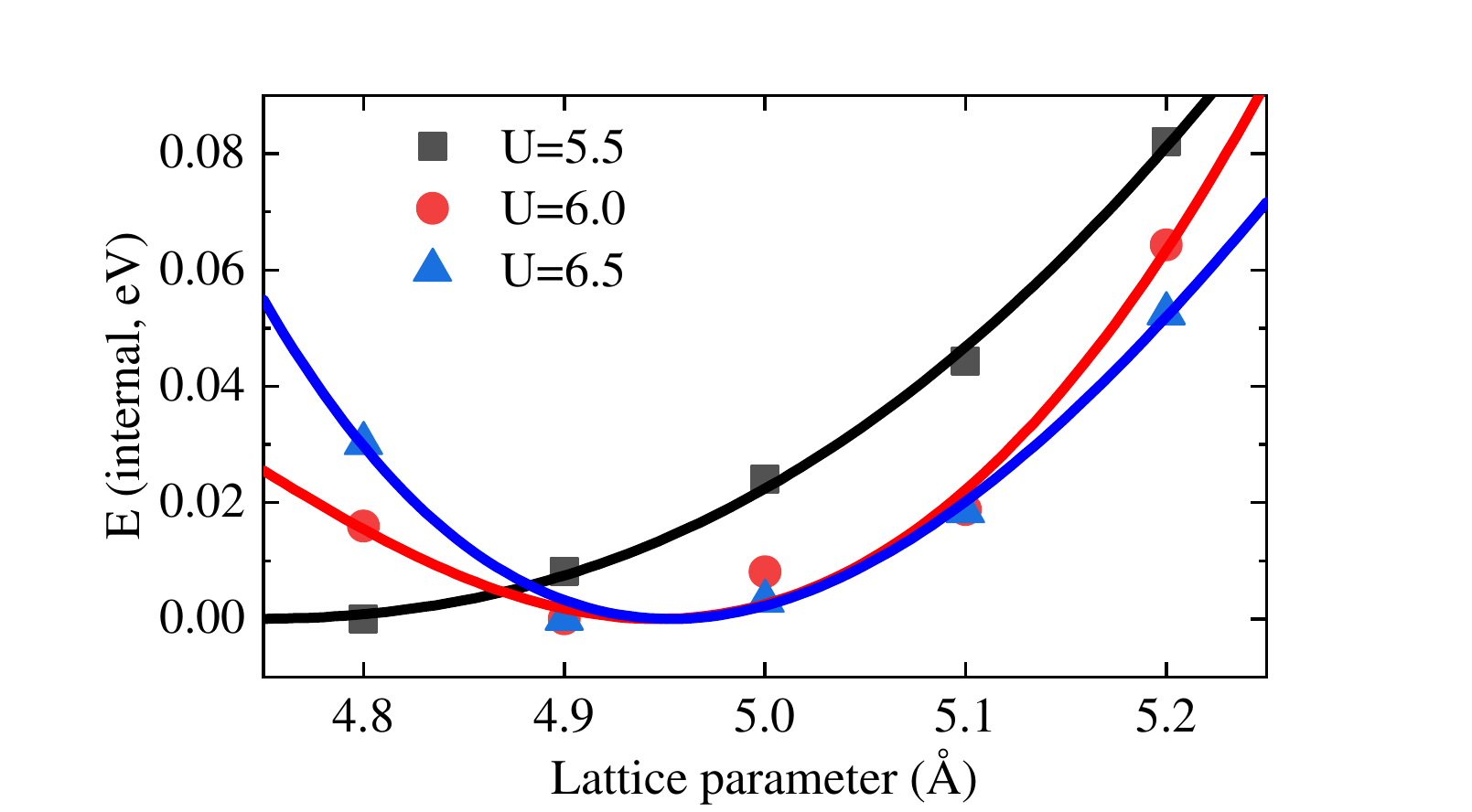}
\includegraphics[width=0.49\columnwidth]{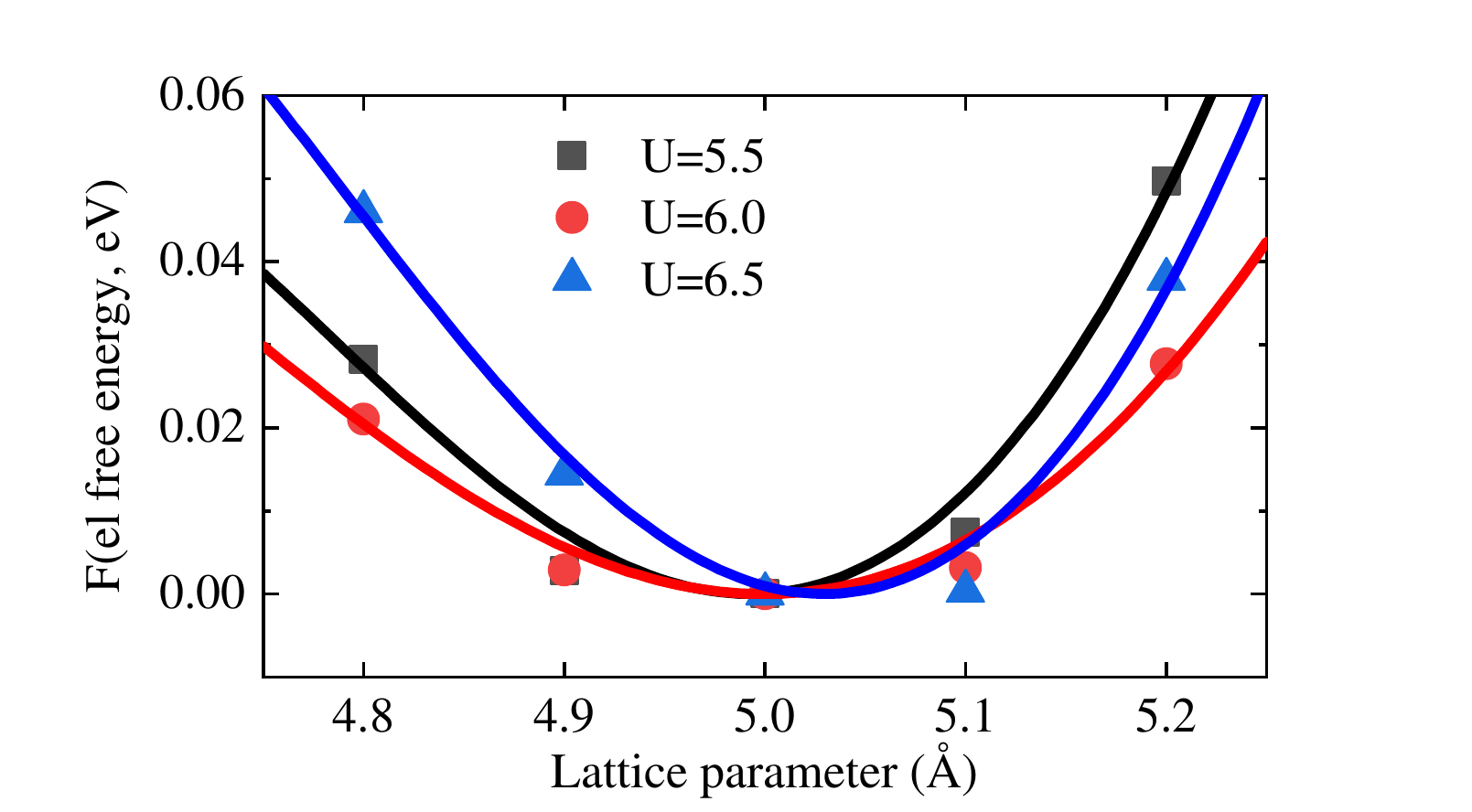}
\caption{LDA+DMFT total energy versus lattice parameter for U = 5.5, 6.0, and 6.5 eV, calculated for lattice constants between 4.8 and 5.2 Å. Symbols denote the calculated data points, while lines represent fits to the Birch–Murnaghan equation of state.}
\label{dmft_u}
\end{figure}

Figure \ref{dmft_u} shows the total energy as a function of the lattice parameter calculated within the LDA+DMFT framework for lattice constants ranging from 4.8 to 5.2 Å, sampled at intervals of 0.1 Å, at Hubbard interaction strengths of U = 5.5, 6.0, and 6.5 eV. For each value of U, a well-defined minimum is observed within this range, allowing the equilibrium lattice parameter to be determined in a controlled manner.

As U increases from 5.5 to 6.5 eV, the position of the energy minimum shifts slightly towards larger lattice parameters, indicating a modest expansion of the equilibrium lattice. The magnitude of this shift remains small and is consistently reproduced across all sampled lattice points, demonstrating that the effect of U on the equilibrium lattice parameter is limited with respect to the discrete lattice sampling.

This weak lattice expansion reflects the enhanced role of electronic correlations within DMFT. With increasing U, the DMFT internal energy and the DMFT electronic free energy become more sensitive to lattice compression, leading to a slightly increased energetic penalty at smaller lattice parameters. As a result, the balance between band energy and correlation energy is marginally shifted in favour of a larger equilibrium lattice at stronger electronic correlations.

\section{S3. Validation of the Fixed Self-Energy Approximation in DMFT Phonon Calculations}

\begin{figure}[ht]
\centering
\includegraphics[width=0.5\columnwidth]{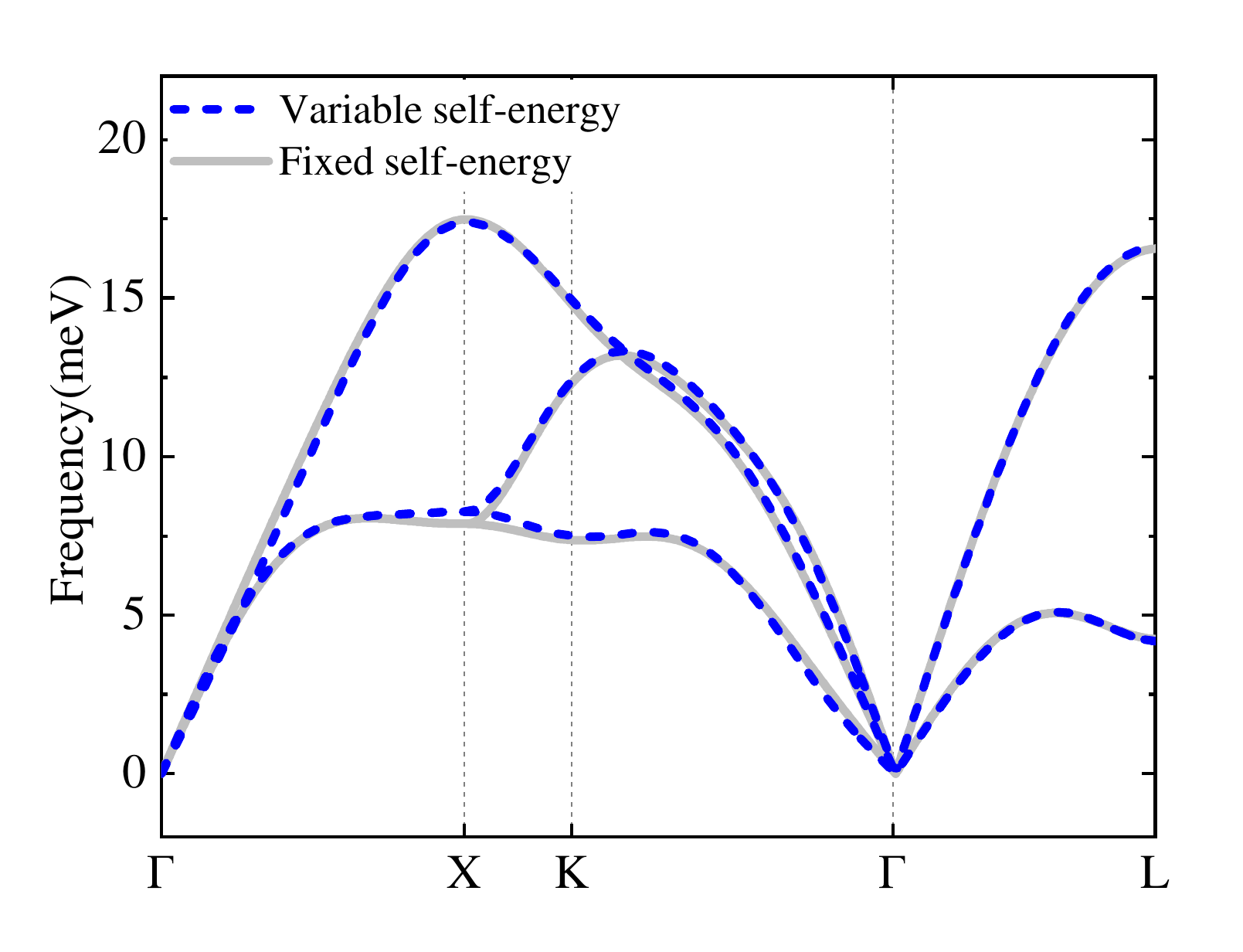}
\caption{Comparison of variable ((full self-consistent)) and fixed self-energy schemes in DMFT-based phonon calculations for $\alpha$-Ce using a $4 \times 4 \times 4$ supercell.}
\label{Variable_Sigma_vs_Fixed_Sigma}
\end{figure}

To evaluate the validity of the fixed self-energy approximation in DMFT-based phonon calculations for cerium, we carried out a direct comparison between variable (full self-consistent) and fixed self-energy schemes for the $\alpha$ phase. The tests were performed at a lattice parameter of 4.799 Å using a $4 \times 4 \times 4$ supercell within the DMFT framework.

As shown in Fig.~\ref{Variable_Sigma_vs_Fixed_Sigma}, the phonon spectra obtained from the two approaches are in excellent agreement, with any differences remaining negligible within numerical accuracy. This result indicates that the lattice dynamical properties of $\alpha$-Ce are only weakly affected by the self-energy variations induced by the small atomic displacements involved in the phonon calculations.

These findings validate the use of the fixed self-energy approximation for the present system. Moreover, this approximation is particularly advantageous for larger supercells, where fully self-consistent DFT+DMFT phonon calculations would become computationally prohibitive.

\clearpage

\section{S4. The influence of SOC on the DFT+DMFT density of states}

\begin{figure}[ht]
\centering
\includegraphics[width=0.6\columnwidth]{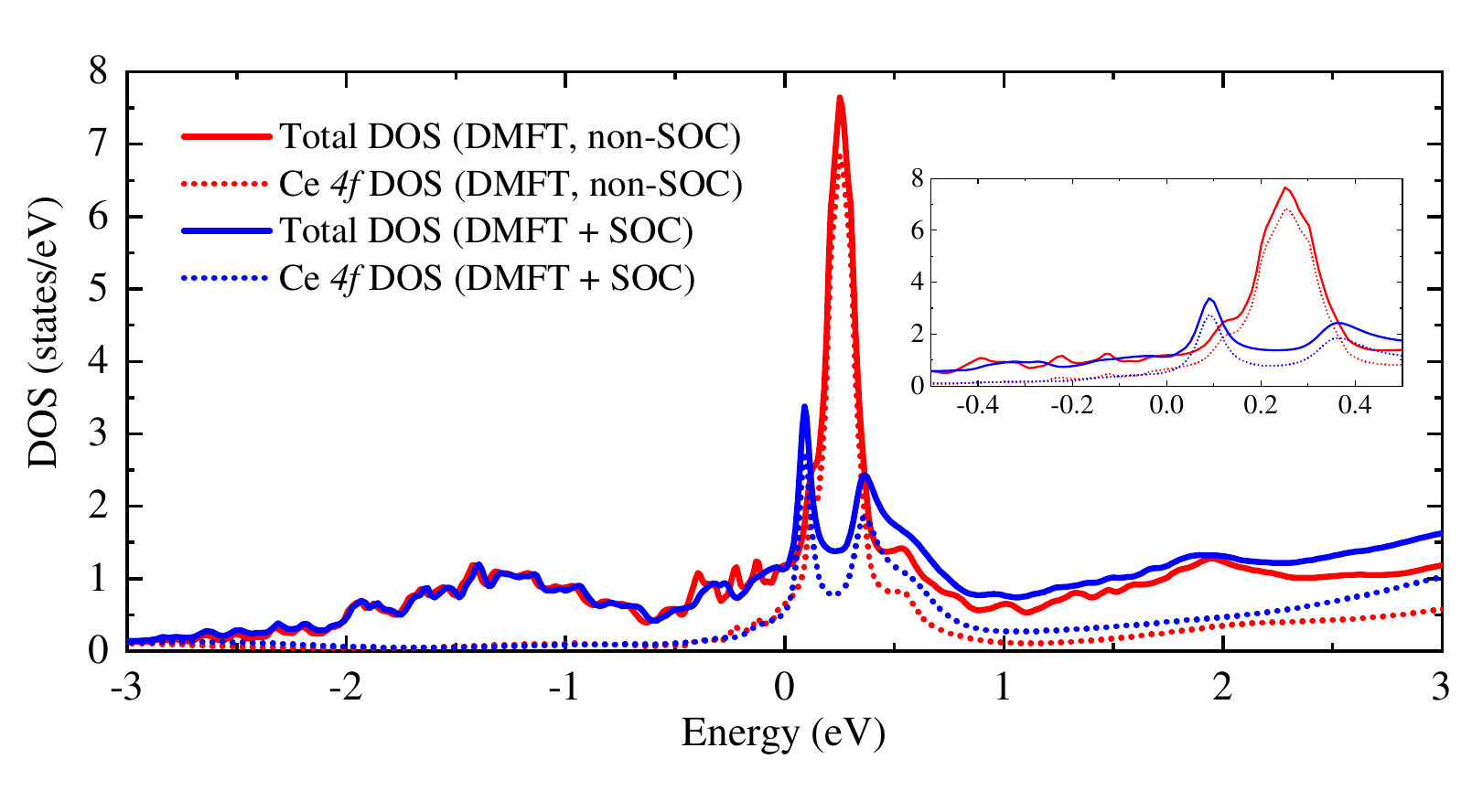}
\caption{Density of states of elemental cerium computed within the LDA+DMFT framework, with and without spin–orbit coupling, at a lattice parameter of 4.799 Å. All energies are referenced to the Fermi level, set to zero. The inset shows an enlarged view of the low-energy region in the vicinity of the Fermi level.}
\label{dos_compare}
\end{figure}

Figure \ref{dos_compare} shows the LDA+DMFT density of states of elemental cerium calculated with and without SOC at a lattice parameter of 4.799 Å. The position and shape of the sharp quasiparticle peak at the Fermi level, as well as its broadening and suppression upon inclusion of SOC, are in excellent agreement with the total and 4f-projected spectral functions reported in the recent configuration-interaction study of the Anderson impurity model for elemental Ce. Without SOC, a sharp quasiparticle peak with dominant Ce 4f character is observed at the Fermi level. When SOC is included, this peak becomes less pronounced and more broadened, reflecting an enhanced many-body renormalisation and a redistribution of low-energy spectral weight. At higher binding energies, the overall density of states remains qualitatively similar \cite{herzog2025configuration}.

\section{S5. The SOC effect on the phonon free energy on the DFT level}

\begin{figure}[ht]
\centering
\includegraphics[width=0.6\columnwidth]{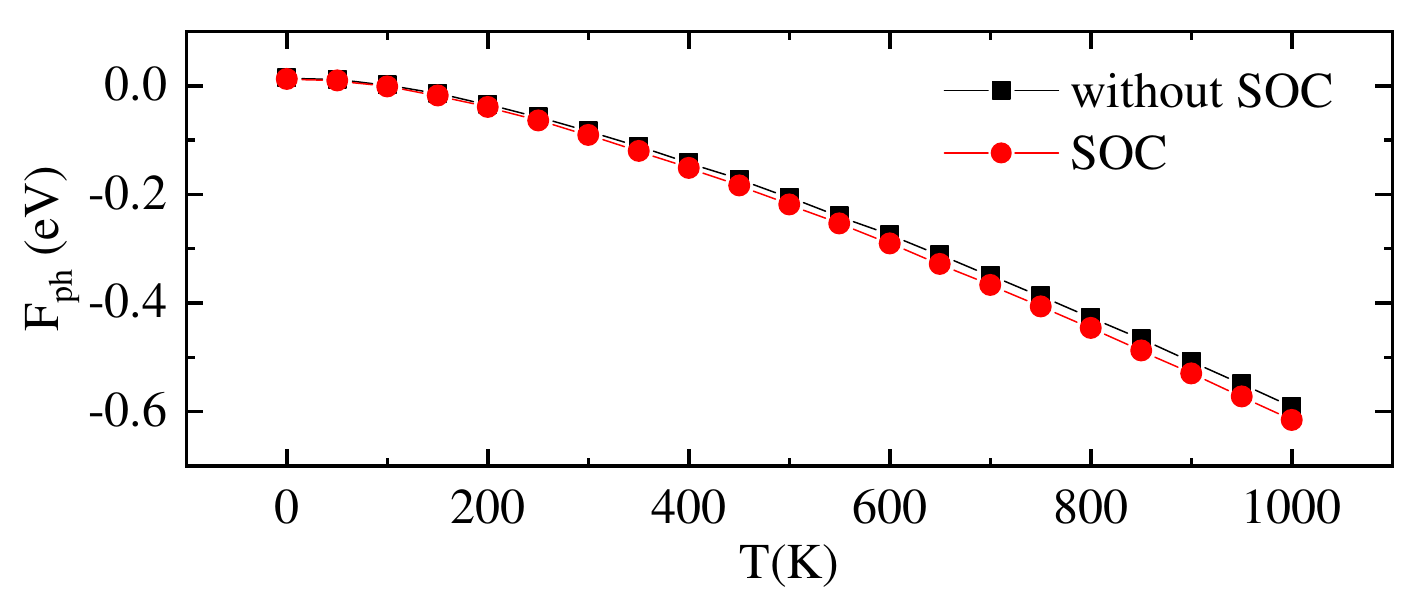}
\caption{
Phonon free energy of cerium as a function of temperature (0--1000~K), calculated within the LDA at a fixed lattice constant of 4.799~\AA, with and without spin--orbit coupling.}
\label{dft_soc_free_energy_compare}
\end{figure}

Figure \ref{dft_soc_free_energy_compare} shows the temperature dependence of the phonon free energy calculated within the LDA at a fixed lattice constant of 4.799~\AA, comparing results obtained with and without spin--orbit coupling. In both cases, the phonon free energy decreases monotonically with increasing temperature. The inclusion of spin--orbit coupling systematically lowers the phonon free energy over the entire temperature range, with the difference becoming more pronounced at elevated temperatures, indicating a non-negligible impact of SOC on the lattice dynamics of cerium.

From the numerical comparison, it is observed that between approximately 500~K and 1000~K the phonon free energy, including spin--orbit coupling, is lower by 5\% in magnitude compared with the calculation without spin--orbit coupling. Based on this analysis, the phonon free energy obtained from DFT at $T \simeq 580$~K is of the order of sub-eV, implying that the additional SOC-induced correction to the phonon free energy is on the order of a few hundredths of an electronvolt. Such a contribution is therefore negligible when compared with the much larger electronic free energy and internal energy terms.

\section{S6. The influence of SOC on the lattice parameter and total energy at the LDA level}

\begin{figure}[h]
\includegraphics[width=0.6\columnwidth]{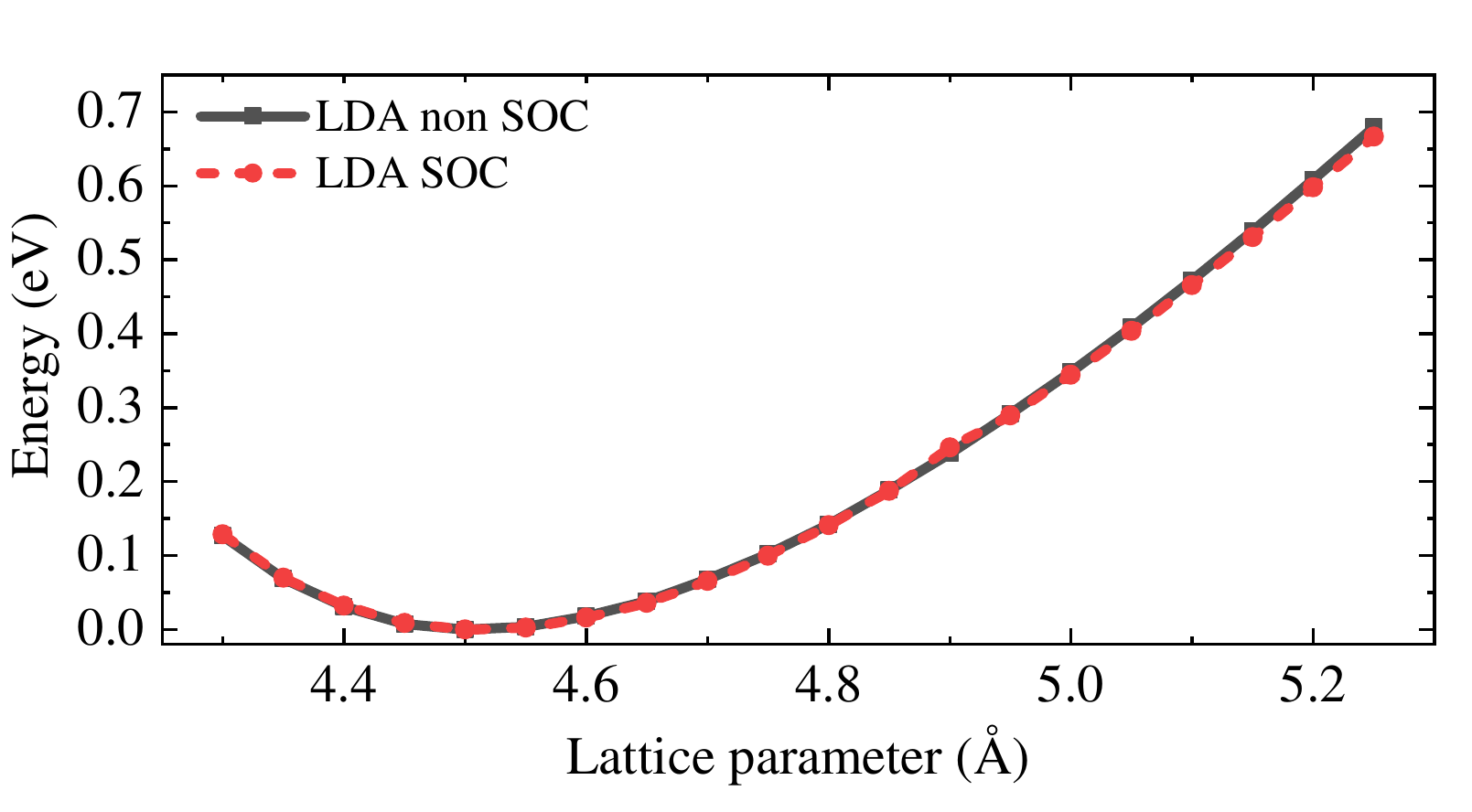}
\caption{Effect of SOC on the LDA Total Energy as a Function of Lattice Parameter}
\label{lda_and_lda_soc_lattice}
\end{figure}

Figure \ref{lda_and_lda_soc_lattice} compares the LDA total energy as a function of the lattice parameter, calculated both with and without SOC. The two energy curves are virtually indistinguishable across the entire volume range considered, yielding identical equilibrium lattice constants and highly similar curvatures. As established in Section S5, the physical difference between these configurations is predominantly the entropic contribution. The absolute energy difference remains exceptionally small. Consequently, the inclusion of SOC exerts a negligible influence on the LDA equation of state, confirming that it plays no significant role in the ground state energetics.

\section{S7. The SOC effect on lattice parameter and electronic free energy at the DMFT level}
\begin{figure}[h]
\includegraphics[width=0.5\columnwidth]{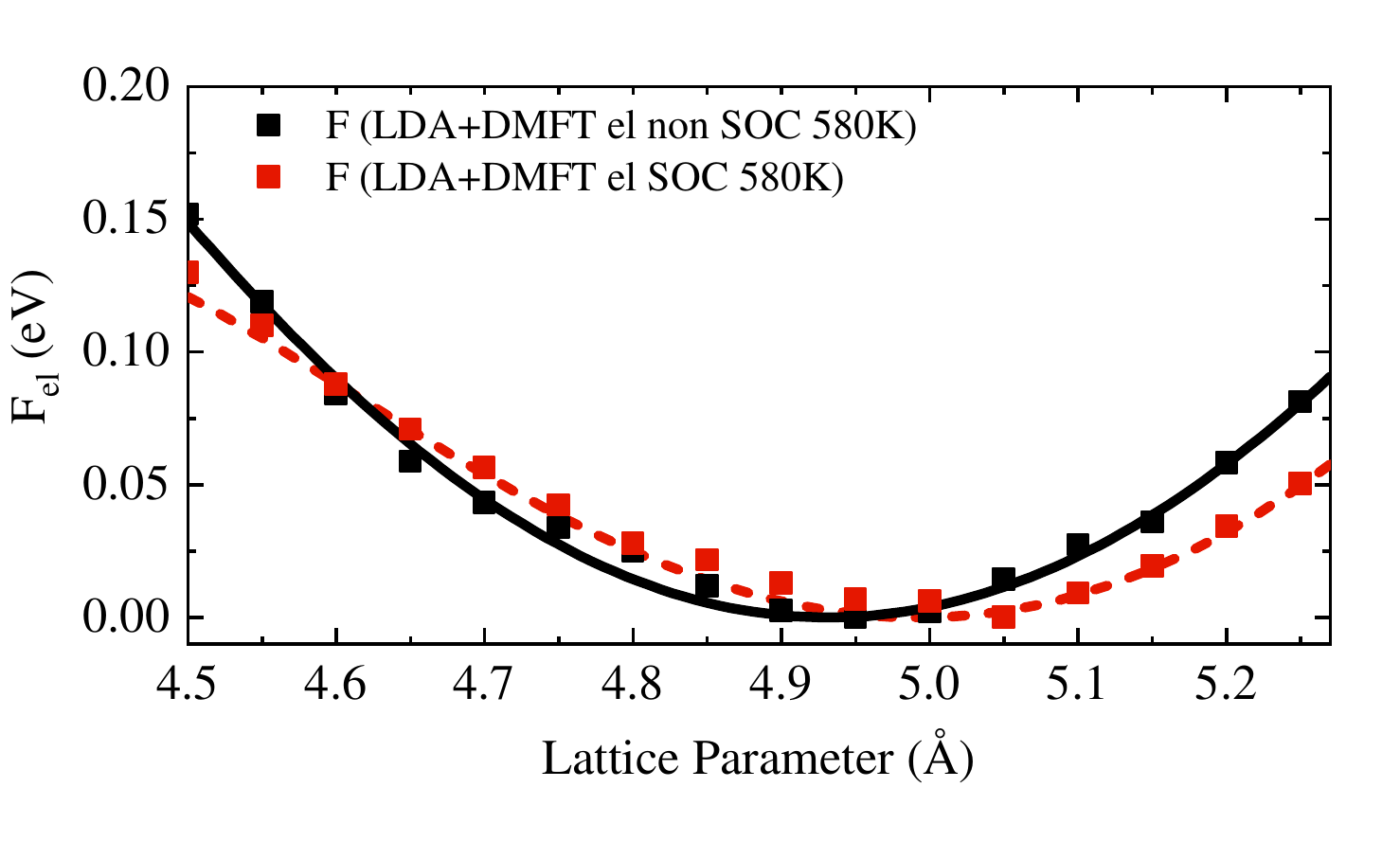}
\caption{Electronic free energy as a function of lattice parameter at 580 K calculated within LDA+DMFT. The red symbols include SOC corrections, while the black symbols correspond to non-SOC results. Symbols denote the calculated data points, while solid lines represent Birch--Murnaghan fits.}
\label{F_el_nsoc_soc}
\end{figure}

\begin{figure}[h]
\includegraphics[width=0.5\columnwidth]{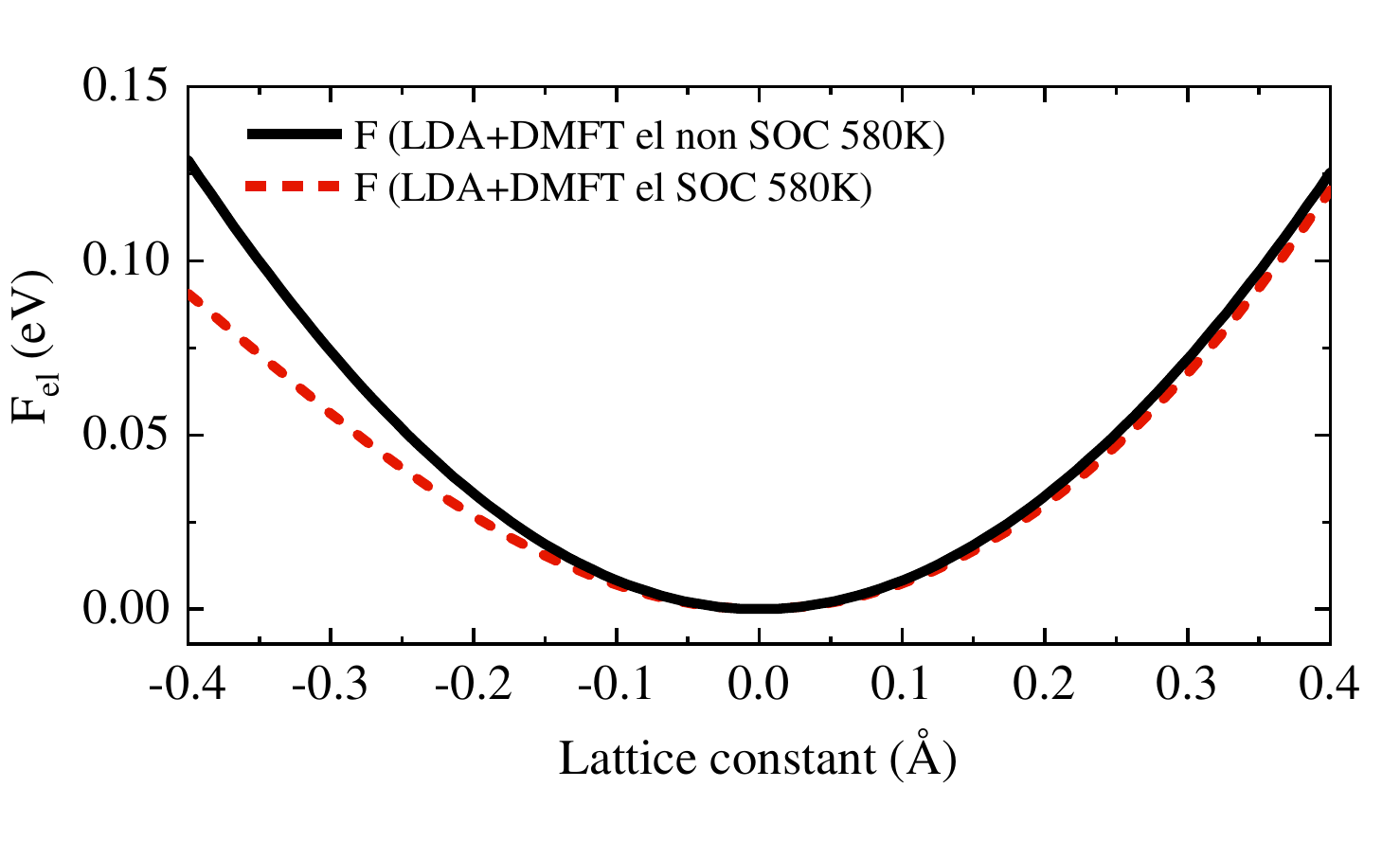}
{\caption{Electronic free energy $F_\mathrm{el}$ (eV) as a function of lattice constant displacement (\AA) at 580~K, obtained from LDA+DMFT calculations without (black solid) and with (red dashed) spin-orbit coupling. The two curves have been rigidly shifted along the displacement axis to align their minima at zero; since such a shift preserves the shape of each curve, the comparison isolates the curvature difference, which quantifies how SOC modifies the electronic free-energy curvature and hence its contribution to the phonon free energy.}}
\label{soc_for_reply}
\end{figure}

Figure \ref{F_el_nsoc_soc} shows the electronic free energy as a function of lattice parameter at 580~K within LDA+DMFT, comparing results with and without SOC. 
{The inclusion of SOC leads to a slight lattice expansion, but the curvature appears very similar.
%
To isolate the effect of SOC on the phonon-relevant curvature, Fig.~\ref{soc_for_reply} replots the same data as a function of displacement from the respective equilibrium lattice parameters. At small displacements the two curves are nearly indistinguishable, demonstrating that, although SOC slightly shifts the equilibrium lattice parameter, the curvature near equilibrium---and hence the restoring forces and phonon frequencies---remains largely unaffected. Deviations appear only at larger displacements, where the anharmonic regions of the two free-energy curves differ.}

{\section{S8.Equilibrium lattice parameter within GGA+DMFT+SOC as a function of U}}

\begin{figure}[ht]
\centering
\includegraphics[width=0.5\columnwidth]{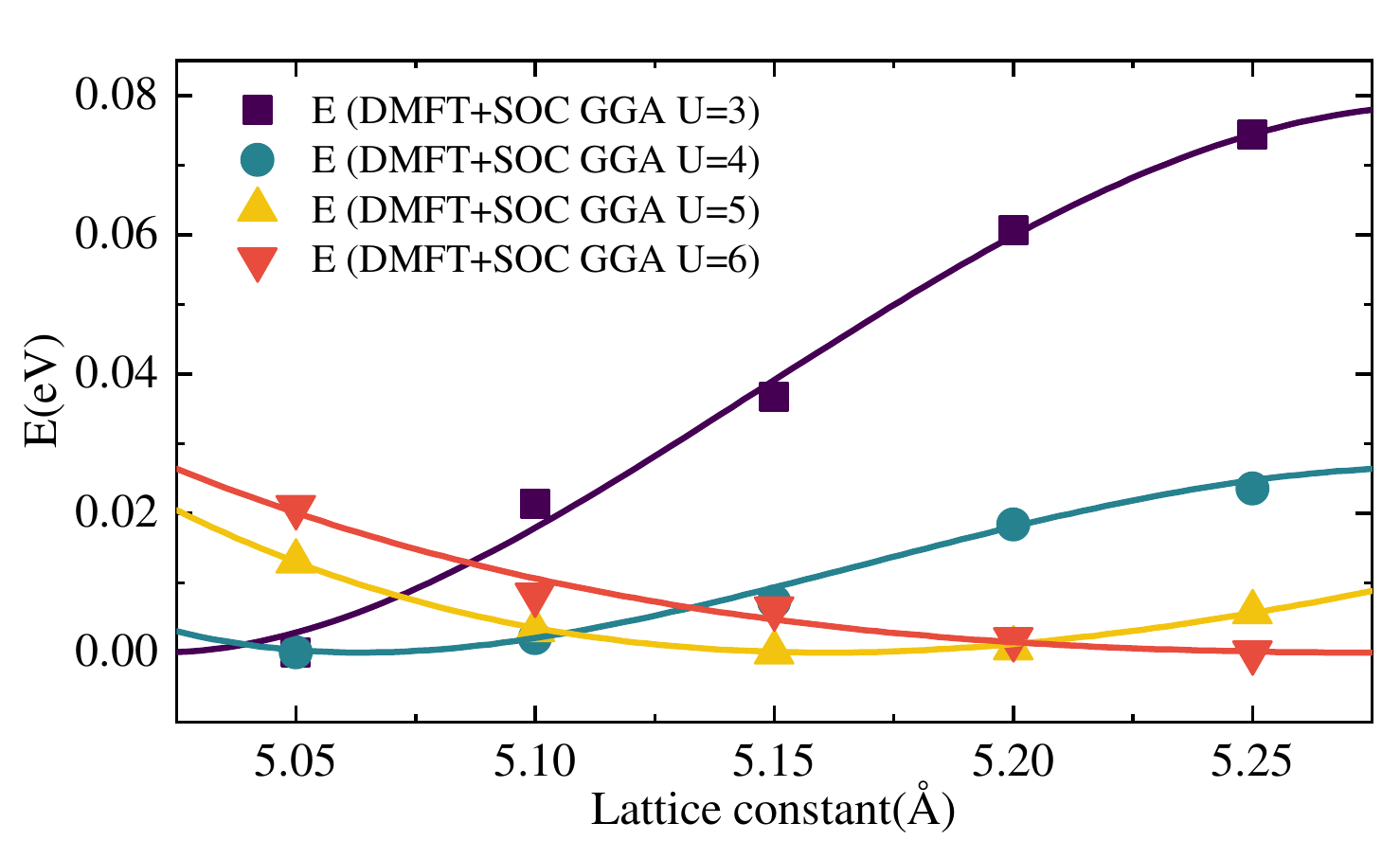}
{\caption{Internal energy of $\gamma$-cerium as a function of lattice constant within GGA+DMFT+SOC at $T = 580$~K, computed for Hubbard interaction strengths $U = 3$, $4$, $5$, and $6$~eV. Symbols denote the calculated data points, while solid lines represent third-order Birch--Murnaghan fits. All curves have their minima aligned to zero. The equilibrium lattice constant is highly sensitive to $U$: the minimum lies below $5.05$~\AA{} for $U = 3$ and $4$~eV and above $5.25$~\AA{} for $U = 6$~eV, while only for $U = 5$~eV does it fall within the sampled window, at approximately $5.15$~\AA{}, close to the experimental $\gamma$-phase value of $5.16$~\AA{}.}}
\label{pbe_dmft_soc_u_lattice}
\end{figure}

{To assess whether the overestimation of the equilibrium lattice constant within GGA+DMFT+SOC ($a_0 = 5.28$~\AA{} at $U = 6$~eV in the main text) can be compensated by reducing the Hubbard interaction, we computed the internal energy as a function of lattice constant at the GGA+DMFT+SOC level for $U = 3$, $4$, $5$, and $6$~eV. Lattice constants were sampled in the range $5.05$--$5.25$~\AA{}, chosen to bracket the experimental $\gamma$-phase value of $5.16$~\AA{}, and the energy curves were fitted to a third-order Birch--Murnaghan equation of state.}

{As shown in Fig.~\ref{pbe_dmft_soc_u_lattice}, the position of the energy minimum depends strongly on $U$. For $U = 3$ and $4$~eV the minimum lies below $5.05$~\AA{}, while for $U = 6$~eV it lies above $5.25$~\AA{}. Only for $U = 5$~eV does the minimum fall within the sampled window, at approximately $5.15$~\AA{}, in close agreement with the experimental value of $5.16$~\AA{}. At the level of the internal energy, $U = 5$~eV therefore provides the best match to experiment within GGA+DMFT+SOC. This result demonstrates two points. First, a smaller U can indeed bring the GGA+DMFT+SOC lattice constant into agreement with experiment at the level of the internal energy. Second, and more importantly, the equilibrium lattice constant is extremely sensitive to U: a change of only 1 eV shifts the minimum across the entire 0.2 Å sampling window. This pronounced sensitivity highlights the lack of predictive power of any approach that relies on tuning U to reproduce the experimental volume.}

\vspace{1em}

\section{S9. Machine-learning-based interpolation of DFT and DFT+DMFT phonon dispersions}

\begin{figure}[ht]
\centering
\includegraphics[width=0.45\columnwidth]{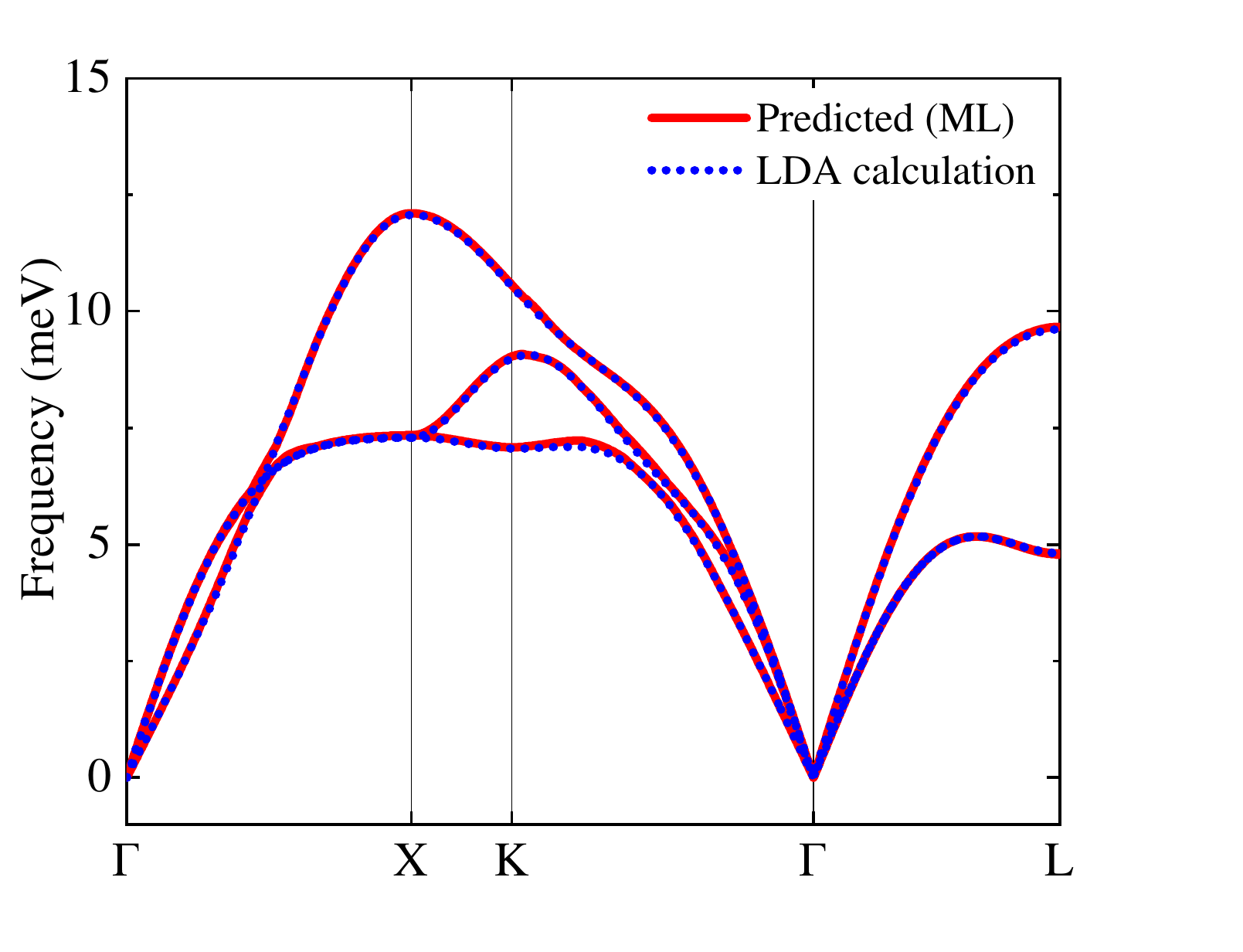}
\includegraphics[width=0.45\columnwidth]{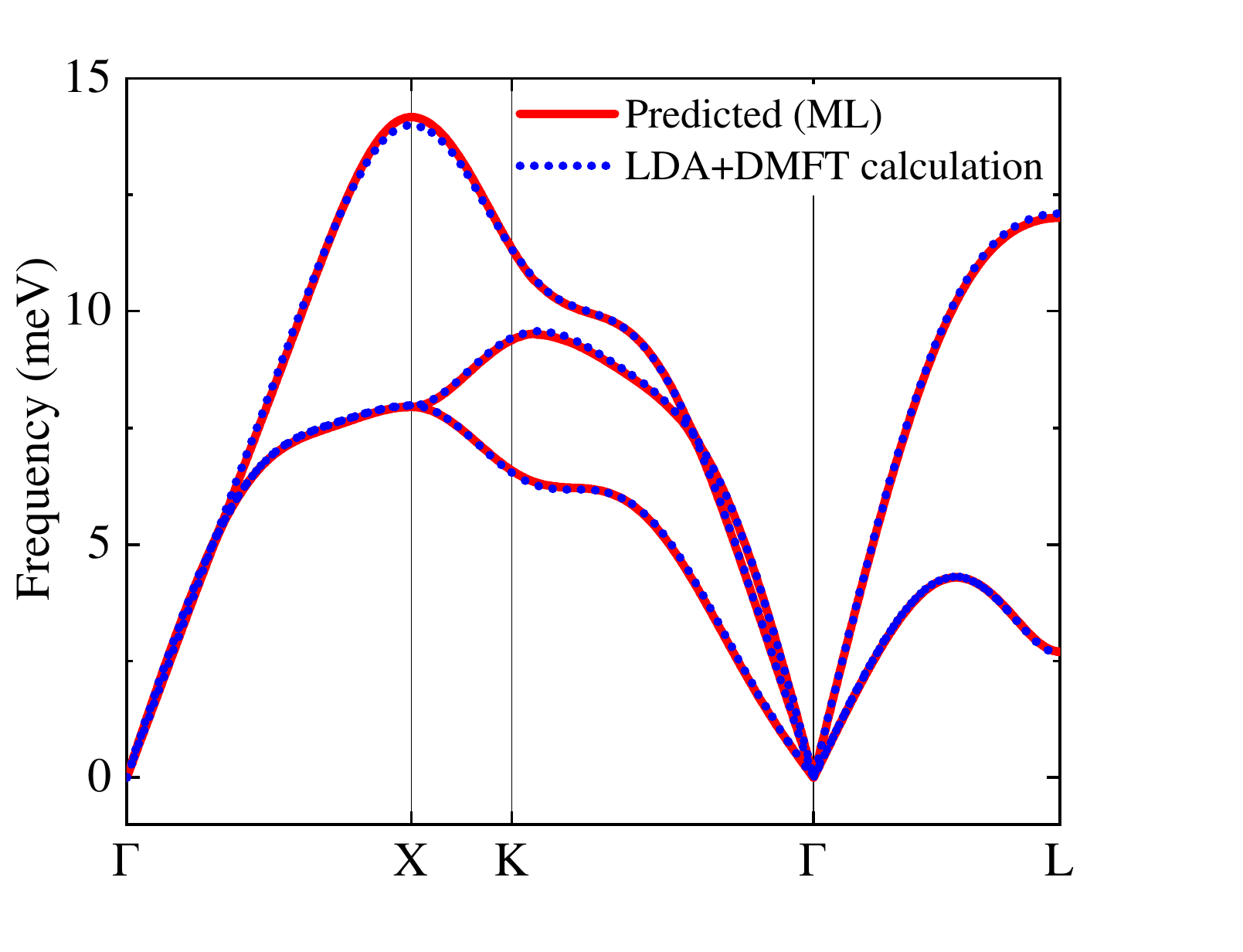}
\caption{Validation of machine-learning-based phonon dispersion interpolation for LDA (left) and LDA+DMFT (right).}
\label{ml}
\end{figure}

Phonon dispersions were computed for a discrete set of lattice constants $a_i$ spanning the range from 4.30~\AA\ to 5.25~\AA\ within both DFT and DFT+DMFT. For all lattice constants, phonon frequencies $\omega(\mathbf{q},\nu;a_i)$ were evaluated along an identical high-symmetry path, using the same set of $\mathbf{q}$ points and phonon branches. Each phonon dispersion can therefore be represented as a single vector,
\begin{equation}
\mathbf{y}(a_i) = (\omega_1,\omega_2,\ldots,\omega_{2997}),
\end{equation}
corresponding to 999 wave vectors and three phonon branches.

Principal component analysis (PCA) \cite{jolliffe2011principal,greenacre2022principal} was applied to the collection of phonon vectors after subtraction of the mean dispersion $\bar{\mathbf{y}}$. Within this framework, a phonon dispersion at an arbitrary lattice constant can be approximated as
\begin{equation}
\mathbf{y}(a) \approx \bar{\mathbf{y}} + \sum_{k=1}^{K} c_k(a)\,\boldsymbol{\Phi}_k,
\end{equation}
where $\boldsymbol{\Phi}_k$ denote lattice-independent principal components and $c_k(a)$ are lattice-dependent projection coefficients. The phonon dispersions are found to admit an efficient low-dimensional representation, with the leading principal component accounting for approximately 98~per~cent of the total variance, and a small number of components sufficient to achieve numerical accuracy.

To model the smooth dependence of the projection coefficients on the lattice parameter, a multilayer perceptron (MLP) neural network \cite{rumelhart1986learning} was employed to interpolate the functions $c_k(a)$. The network comprises multiple hidden layers with nonlinear activation functions, which are adequate to capture the nonlinear evolution of the coefficients under lattice expansion or compression. The phonon dispersion at any intermediate lattice constant is then reconstructed through the inverse PCA transformation.

The accuracy of this interpolation scheme was assessed using a leave-one-out validation, as shown in Fig.~\ref{ml}. In this test, the phonon dispersion at $a=5.00$~\AA\ was excluded from the dataset, and the model was trained using the remaining lattice constants. The resulting prediction at $a=5.00$~\AA\ is in excellent agreement with the directly calculated phonon dispersions obtained from both DFT and DFT+DMFT, across all phonon branches and high-symmetry points. As the test lattice constant lies within the training interval, this comparison demonstrates genuine interpolation and confirms that the method faithfully captures the smooth, collective evolution of the phonon spectrum with lattice parameter.

\begin{figure}[ht]
\centering
\includegraphics[width=0.6\columnwidth]{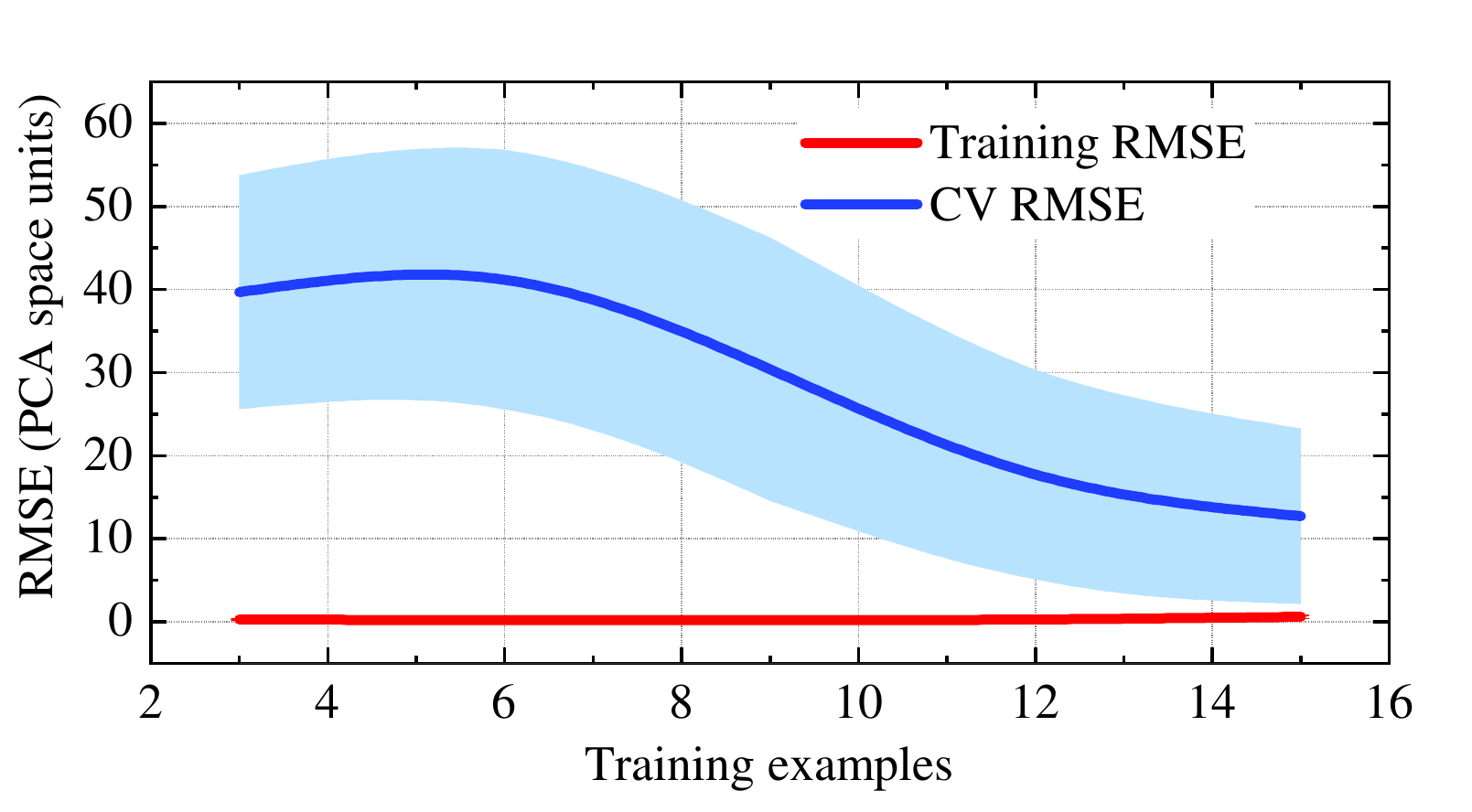}
\caption{Learning curve of the PCA-MLP model.
The root-mean-square error (RMSE) in PCA space is plotted against the training set size. }
\label{ML_training}
\end{figure}

To validate the efficiency and accuracy of our interpolation scheme specifically for the computationally demanding LDA+DMFT phonon spectra, we performed a systematic learning curve analysis, as shown in Fig. \ref{ML_training}. The training error (red line) remains consistently near zero, indicating that the network possesses sufficient capacity to capture the full complexity of the many-body renormalised force constants without information loss. Simultaneously, the cross-validation error (blue line) decreases sharply as the number of training samples increases, accompanied by a significant narrowing of the prediction variance (shaded region). This convergence behaviour confirms that the collective softening of the phonon spectrum driven by electronic correlations is smooth and low-dimensional. Consequently, the dataset of 15 lattice configurations calculated via LDA+DMFT is sufficient to achieve a robust and high-fidelity representation of the phonon dispersions across the entire volume range of the $\alpha-\gamma$ transition.

\section{S10. Temperature-dependent total free-energy analysis at the DFT level}

\begin{figure}[h]
\includegraphics[width=0.6\columnwidth]{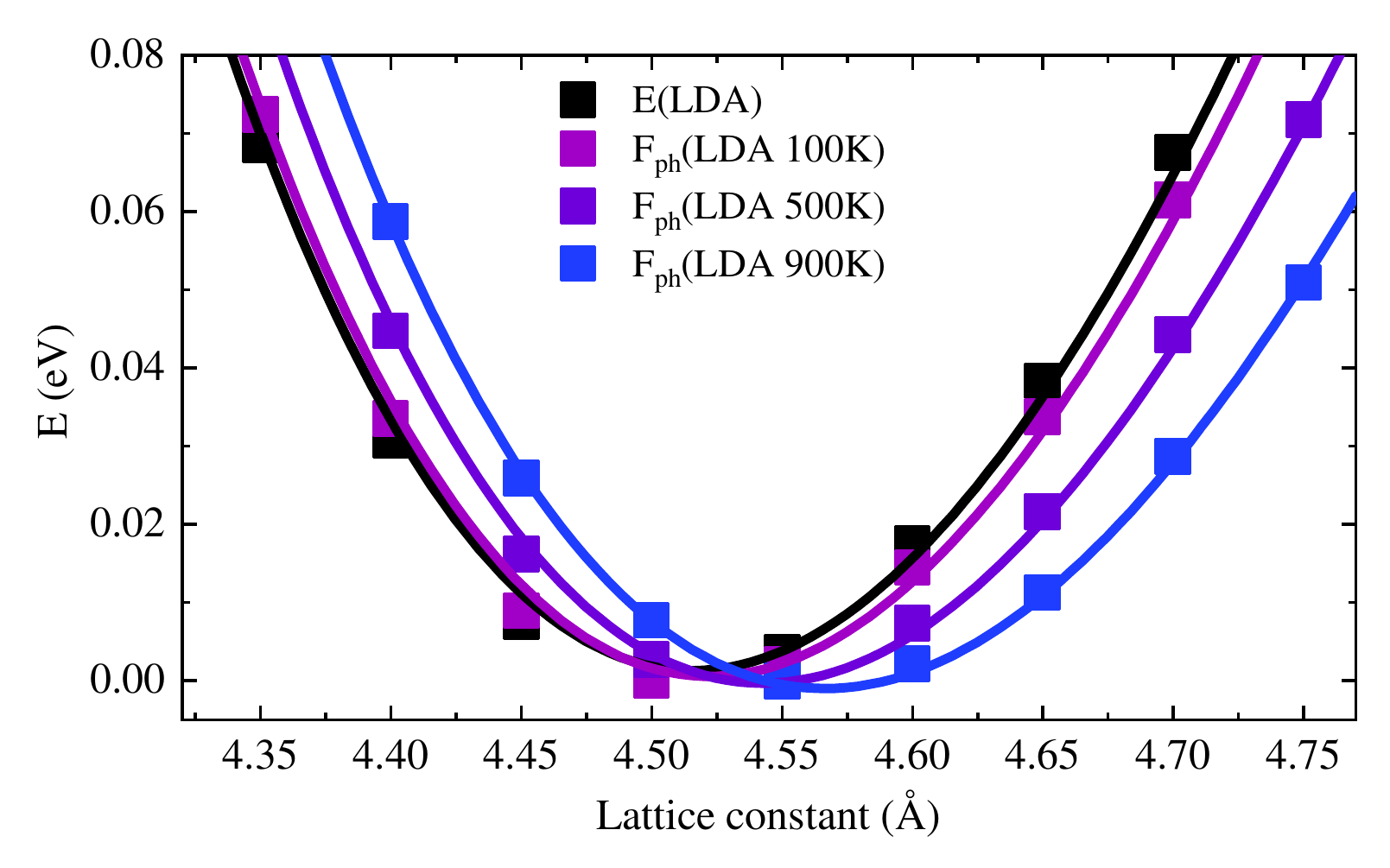}
\caption{Total energy of cerium as a function of the lattice constant calculated within DFT, corrected by the phonon free energy. Symbols denote the calculated data points, while solid lines represent Birch--Murnaghan fits.}
\label{lda_free_energy}
\end{figure}

Figure~\ref{lda_free_energy} shows the total energy of fcc cerium as a function of the lattice constant after incorporating the phonon free-energy correction within the DFT framework. The phonon free energy, obtained from the vibrational density of states, is added to the static DFT total energy to construct a temperature-dependent free-energy landscape. The resulting energy curves are fitted using the third-order Birch--Murnaghan equation of state, from which the equilibrium lattice constants are extracted at each temperature.

In the absence of phonon corrections, the equilibrium lattice constant is found to be $a_0 = 4.52$~\AA. Upon including the vibrational free energy, the minimum of the free-energy curve shifts systematically towards larger lattice constants with increasing temperature, yielding $a_0 = 4.52$~\AA\ at 100~K, which is nearly identical to the uncorrected value. Moreover it is $a_0 = 4.54$~\AA\ at 500~K, and $a_0 = 4.57$~\AA\ at 900~K. This monotonic lattice expansion reflects the increasing contribution of vibrational entropy, which progressively stabilises larger volumes at elevated temperatures.

The temperatures 100, 500, and 900~K are chosen to span the experimentally relevant regime in which cerium remains in the face-centred cubic structure at ambient pressure, ensuring that the phonon free-energy corrections are evaluated consistently within a single crystallographic phase~\cite{krisch2011phonons}. These results demonstrate that even at the DFT level, lattice vibrational effects provide a non-negligible renormalisation of the equilibrium volume of cerium and must be taken into account for a quantitative description of its thermodynamic properties.

\section{S11. Temperature dependence of the DMFT internal energy and electronic free energy}

\begin{figure}[h]
\centering
\includegraphics[width=0.5\columnwidth]{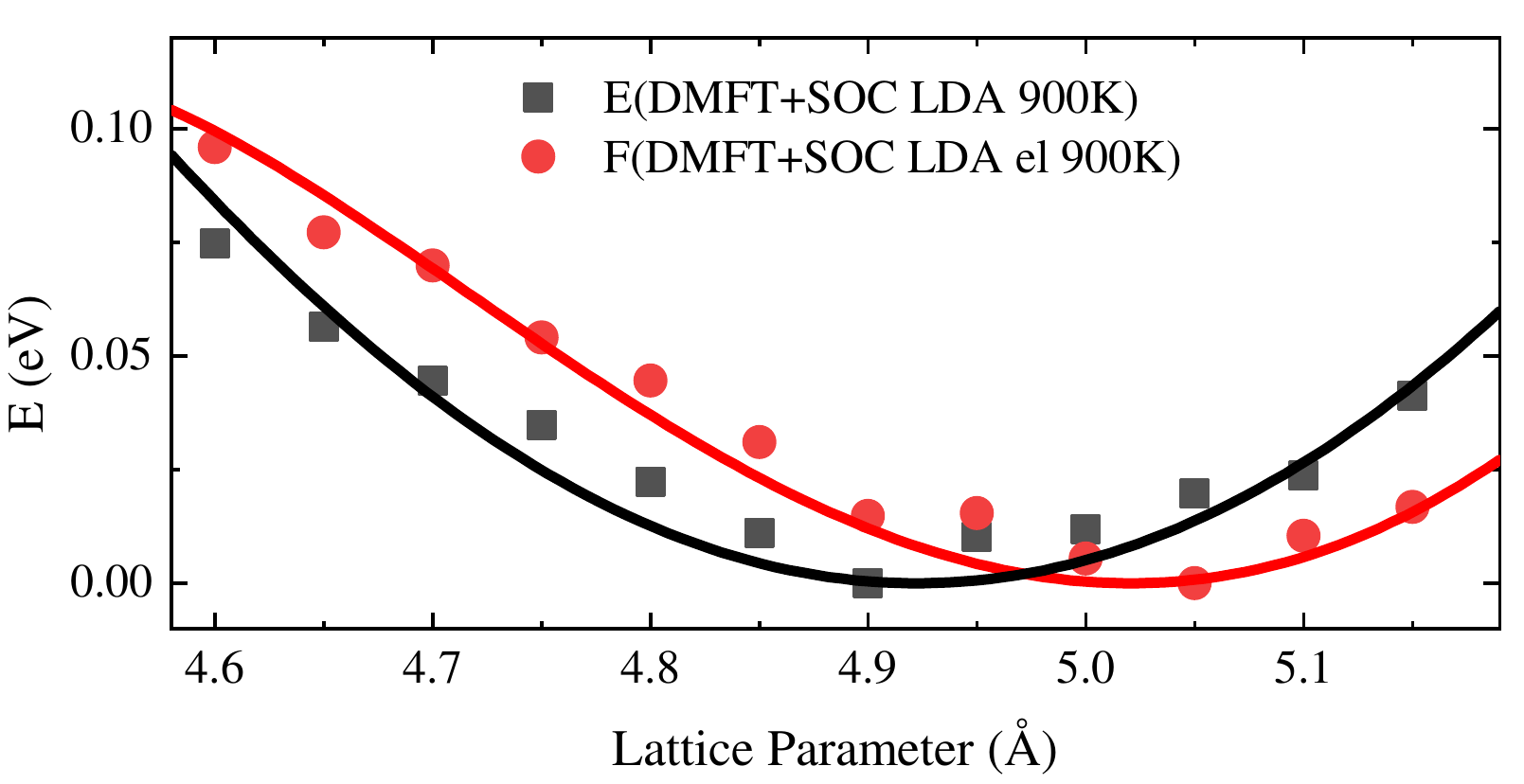}
\caption{Temperature dependence of the DMFT internal energy and electronic free energy at 900 K. Symbols denote the calculated data points, while solid lines represent Birch--Murnaghan fits.}
\label{dmft_900K}
\end{figure}

Figure~\ref{dmft_900K} shows the lattice-parameter dependence of the internal energy and the electronic free energy, including SOC, obtained within the LDA+DMFT framework at 900~K. The positions of the minima are very close to those obtained at 580~K in the main text, with only a marginal lattice expansion of approximately 0.03--0.04~\AA.

\bibliography{apssamp}